\documentclass[12pt,preprint]{aastex}
\usepackage{geometry}               
\usepackage{amssymb}
\usepackage{amsmath}
\usepackage{graphicx}
\usepackage{color}
\usepackage{txfonts}

\usepackage{fancyhdr}
\usepackage{wrapfig}

\def\beq{\begin{equation}}
\def\eeq{\end{equation}}

\begin{document}

\shorttitle{Radiative Magnetic Reconnection}
\shortauthors{D. A. Uzdensky}

\title{Radiative Magnetic Reconnection in Astrophysics}

\author{Dmitri A. Uzdensky}
\affil{Center for Integrated Plasma Studies, Physics Department, University of Colorado, Boulder, CO 80309 \email{uzdensky@colorado.edu}}

\date{August 31, 2015}

\maketitle


{\bf Abstract:}
In this Chapter we review a new and rapidly growing area of research in high-energy plasma astrophysics --- radiative magnetic reconnection, defined here as a regime of reconnection where radiation reaction has an important influence on the reconnection dynamics, energetics, and/or nonthermal particle acceleration. 
This influence be may be manifested via a variety of radiative effects that are critical in many high-energy astrophysical applications. The most notable radiative effects in astrophysical reconnection include radiation-reaction limits on particle acceleration, radiative cooling, radiative resistivity, braking of reconnection outflows by radiation drag, radiation pressure, viscosity, and even pair creation at highest energy densities.   
The self-consistent inclusion of these effects into magnetic reconnection theory and modeling sometimes calls for serious modifications to our overall theoretical approach to the problem. 
In addition, prompt reconnection-powered radiation often represents our only observational diagnostic tool available for studying remote astrophysical systems; this underscores the importance of developing predictive modeling capabilities to connect the underlying physical conditions in a reconnecting system to observable radiative signatures. 
This Chapter presents an overview of our recent theoretical progress in developing basic physical understanding of radiative magnetic reconnection, with a special emphasis on astrophysically most important radiation mechanisms like synchrotron, curvature, and inverse-Compton. The Chapter also offers a broad review of key high-energy astrophysical applications of radiative reconnection, illustrated by multiple examples such as: pulsar wind nebulae, pulsar magnetospheres, black-hole accretion-disk coronae and hot accretion flows in X-ray Binaries and Active Galactic Nuclei and their relativistic jets, magnetospheres of magnetars, and Gamma-Ray Bursts. 
Finally, this Chapter discusses the most critical open questions and outlines the directions for future research of this exciting new frontier of magnetic reconnection research.

\newpage
\section{Introduction}
\label{sec-intro}

Magnetic reconnection --- a process of rapid rearrangement of magnetic field lines corresponding to a change of magnetic field topology --- is widely recognized as one of the most important fundamental plasma physical processes \citep{Biskamp-2000, Zweibel_Yamada-2009, Yamada_etal-2010}.  It is ubiquitous in laboratory, space, and astrophysical plasmas. As with many other important plasma processes, the reason why it is considered to be important and why researchers are interested in it, is that it is one of the processes controlling the plasma energetics, i.e., energy exchange between the different plasma system constituents. 
The specific case of reconnection involves the flow of energy from the magnetic field to the charged particles comprising the plasma.  
This energy conversion is made  possible by the relaxation of magnetic field to a lower energy state suddenly made accessible by the breaking of topological constraints in reconnection, in particular, the flux freezing constraint.  

Although on the fundamental particle level, the released energy goes to the kinetic energy of individual particles, it is customary to describe the resulting energization of the plasma by splitting it into several parts at the fluid level, such as the thermal heating of plasma and the bulk kinetic energy of the reconnection outflows, plus nonthermal particle acceleration at the kinetic level of plasma description. 
The question of the partitioning of the dissipated magnetic energy among these different forms of plasma energy (as well as the partitioning of energy between electrons and ions) is considered to be one of the main driving questions in magnetic reconnection research (see, e.g., the chapter by Yamada et al. in this volume).  
To a large extent, this is because  it most directly relates to the observed radiation powered by reconnection in remote astrophysical sources. In fact, one of the main reasons why scientists are interested in reconnection, and especially in its energetics aspects, is that this process is commonly believed to be responsible for some of the most spectacular and energetic phenomena observed in the Universe. 
In particular, it is believed to be the mechanism for powering many explosive phenomena exhibiting very rapid time variability and impulsive character --- various kinds of {\it flares}  and bursts of high-energy (UV, X-ray, and gamma-ray) radiation.  

Reconnection is especially important as an energetically dominant mechanism in systems that are magnetically dominated (low plasma-$\beta$), that is in tenuous hot coronae and magnetospheres of dense and relatively cold gravitationally stratified objects~\citep{Uzdensky-2006}. 
The two most prominent and best-studied, classic examples of reconnection in Nature are solar flares and magnetic substorms in the Earth magnetosphere (see, e.g., the chapters by Shibata \& Takasao, by Cassak \& Fuselier,  and by Petrukovich et al. in this volume).
This is the area where reconnection research has started more than 50 years ago \citep{Sweet-1958, Parker-1957, Dungey-1961, Axford-1967, Vasyliunas-1975} and where the case for reconnection is most convincingly established by observations \citep{Masuda_etal-1994, Shibata-1996,  Yokoyama_Shibata-1995, Yokoyama_etal-2001, Tsuneta-1996, Paschmann_etal-2013}. 

However, in the past couple of decades, reconnection has also been increasingly often proposed (although without such a strong observational evidence as in the above heliospheric examples) as an attractive possible mechanism for powerful flares in many astrophysical systems outside the solar system, especially in high-energy astrophysics. 
It has been invoked to explain energy dissipation and radiation in pulsar systems (e.g., in pulsar magnetospheres, the striped pulsar winds, and pulsar wind nebulae, PWNe) \citep{Michel-1982, Coroniti-1990, Michel-1994, Lyubarsky-1996, Lyubarsky-2000, Lyubarsky_Kirk-2001, Kirk_Skjaeraasen-2003, Petri_Lyubarsky-2007, Contopoulos-2007, Lyutikov-2010, Uzdensky_etal-2011, Bednarek_Idec-2011, Cerutti_etal-2012a, Cerutti_etal-2013, Cerutti_etal-2014a, Cerutti_etal-2014b, Sironi_Spitkovsky-2011, Clausen-Brown_Lyutikov-2012, Arka_Dubus-2013, Uzdensky_Spitkovsky-2014, Philippov_etal-2014, Philippov_etal-2015, Cerutti_etal-2015}; in gamma-ray bursts (GRBs) \citep{Spruit_etal-2001, Drenkhahn_Spruit-2002, Lyutikov-2006b, Giannios_Spruit-2007, McKinney_Uzdensky-2012};  magnetospheres of magnetars  \citep{Lyutikov-2003b, Lyutikov-2006a, Masada_etal-2010, Uzdensky-2011, Parfrey_etal-2012, Parfrey_etal-2013}, and in coronae and jets of accreting black holes (BHs)  
\citep{GRV-1979, van_Oss_etal-1993, deGouveia_etal-2005, Uzdensky_Goodman-2008, Goodman_Uzdensky-2008, deGouveia_etal-2010, Khiali_etal-2015a, Kadowaki_etal-2015, Singh_etal-2015}, including those in active galactic nuclei (AGN) and blazars \citep{Romanova_Lovelace-1992, DiMatteo-1998, DiMatteo_etal-1999, Lesch_Birk-1998, Schopper_etal-1998, Larrabee_etal-2003, Liu_etal-2003, Lyutikov-2003a, Jaroschek_etal-2004a, Jaroschek_etal-2004b, Giannios_etal-2009, Giannios_etal-2010, Giannios-2010,  Nalewajko_etal-2011, Khiali_etal-2015b}.

It is worth noting that in most traditional, solar-system applications of reconnection, including solar flares, Earth magnetosphere, and sawtooth crashes in tokamaks, one rightfully ignores radiation emitted promptly during the reconnection process. This is well justified because in these situations the radiative cooling time of the energetic particles is usually much longer than the time they spend in the reconnection region and, in fact, than the entire duration of the reconnection event. 

In contrast, however,  in many high-energy astrophysical environments, especially relativistic ones, the ambient magnetic and radiation energy density is so high and the system size is so large that reconnection takes place in the radiative regime.  
This means that the radiation reaction force on the emitting particles is rather strong and needs to be taken into account because it materially affects the particles' motion.  
This makes it necessary to understand  {\it radiative magnetic reconnection}, which we define here as a reconnection regime in which radiation back-reaction has an important effect on reconnection dynamics, energetics, and/or particle acceleration.  In this regime magnetic dissipation and radiative processes intertwine and influence each other.  Understanding how this happens represents an exciting new frontier in plasma astrophysics. This frontier is only now beginning to being charted and the main goal of this article is to give an overview of the recent progress in this area. 

In principle, radiation reaction can exert several effects on reconnection: 
(1) putting an upper limit on the high-energy extent of nonthermal particle acceleration and hence on the energy of the emitted photons; 
(2) optically-thin or optically-thick radiative cooling; 
(3) radiative drag on the current-carrying electrons in the layer (radiative resistivity); 
(4) radiation drag on the reconnection outflows 
(manifested as an effective radiation viscosity in the optically-thick case); 
(5) radiation pressure; 
(6) pair creation. 
The main radiative mechanisms in high-energy astrophysical plasmas, especially relativistic ones, are cyclotron/synchrotron emission, curvature emission, and inverse-Compton (IC) scattering. 

Apart from its significance to basic plasma physics, the main motivation for exploring radiative reconnection comes from numerous astrophysical applications.  Examples of radiative relativistic reconnection in astrophysics include: 

(1) Accretion flows and accretion disk coronae (ADC) around black holes with accretion rates approaching the Eddington limit, in both stellar-mass galactic X-ray binary (XRB) systems and in supermassive BHs in AGN, e.g., quasars. Here, reconnection processes occur in the presence of a very intense radiation field emitted by the disk, which leads to a very powerful inverse-Compton (IC) cooling of the electrons energized by reconnection \citep{Goodman_Uzdensky-2008, Khiali_etal-2015a, Khiali_etal-2015b}. 

(2) Magnetospheres and relativistic winds of pulsars (rapidly rotating magnetized neutron stars), where reconnection should  happen in the ballerina-skirt equatorial current sheet outside the pulsar light cylinder (LC).  In many cases (e.g., the Crab), the magnetic field at the LC is so high that prompt synchrotron cooling of the plasma heated by reconnection is very strong; it controls the energy balance of the layer and limits the plasma temperature. 
This radiation may then explain the powerful pulsed high-energy (GeV) $\gamma$-ray emission observed in Crab and other pulsars (e.g., \citep{Lyubarsky-1996, Uzdensky_Spitkovsky-2014}, Fish et~al.~2015, in prep.).

(3) Pulsar Wind Nebulae (PWN), including the Crab Nebula; here, although radiative cooling is not strong enough to affect the bulk of the plasma, synchrotron radiation reaction may limit the extreme (PeV) particle acceleration, which has important implications for the $\gamma$-ray flares recently discovered in the Crab Nebula \citep{Bednarek_Idec-2011, Uzdensky_etal-2011, Cerutti_etal-2012a, Cerutti_etal-2013, Cerutti_etal-2014a, Cerutti_etal-2014b}.  

(4) Ultra-relativistic jets in blazars where radiative braking and cooling may alter reconnection dynamics and radiation production, e.g., in the context of very rapid TeV flares in several blazar systems \citep{Giannios_etal-2009, Nalewajko_etal-2011, Giannios-2013}. 

5) Gamma-Ray Bursts (GRBs), where magnetic reconnection has been conjectured to power the main prompt gamma-ray emission \citep{Drenkhahn_Spruit-2002, Giannios_Spruit-2006}.  Here, reconnection takes place in an environment with high energy density and large optical depth, so that photon resistivity, radiation cooling, and radiation pressure need to be taken into account \citep{McKinney_Uzdensky-2012}.

(6) Magnetospheres of magnetars (ultra-magnetized neutron stars), where it has been suggested that, by analogy with the solar corona, reconnection may explain the observed powerful gamma-ray flares (e.g., \citep{Lyutikov-2006a, Uzdensky-2011, Parfrey_etal-2012, Parfrey_etal-2013, Yu-2012}).  The energy density in these systems is extremely high  and hence reconnection inevitably leads to relativistically-hot temperatures and copious photon and pair creation \citep{Uzdensky-2011, Uzdensky_Rightley-2014}. 

One can thus see that the large number and the great diversity of examples of radiative magnetic reconnection in astrophysics strongly motivate advancing this new research frontier.

Another practical reason to study radiative reconnection is that, in astrophysics, 
remote telescopic observations of the radiation produced in a flare provide our only diagnostic probe for studying the underlying physics. 
For this reason, the ability to calculate the observable radiation spectrum is critical for testing whether reconnection (or any other given process) can explain observations.
Thus, in order to connect our theoretical/numerical reconnection models with the actual observable radiation, we must develop a rigorous method for calculating the produced radiation signatures in detail, such as  time-resolved spectra for a given orientation of the observer's line of sight.

Finally, radiative reconnection should also be of potential considerable interest to experimental High-Energy-Density (HED) Physics, a new branch of modern physics that has emerged in recent years. 
One can anticipate rapid progress in HED reconnection studies facilitated by new experimental capabilities developed in the HED Physics community, e.g., made possible by powerful lasers (such as Omega EP and NIF) and Z-pinches (e.g., Magpie \citep{Lebedev_etal-2014, Suttle_etal-2014}). 
In fact, several HED reconnection experiments utilizing laser-produced plasmas with mega-gauss magnetic fields have already been reported (e.g., \citep{Nilson_etal-2006, Li_etal-2007, Nilson_etal-2008, Dong_etal-2012,  Fox_etal-2011, Fox_etal-2012, Fiksel_etal-2014}).

This review Chapter is organized as follows. 
Before embarking on our main discussion of the effects that radiation may exert on reconnection,
in~\S~\ref{sec-passive-rad-signs} we first make some remarks about passive radiative signatures of reconnection (\S\S~\ref{subsec-passive-general}-\ref{subsec-passive-nonthermal}), including applications to hard X-ray bremsstrahlung emission in solar flares (\S~\ref{subsec-loop_top}) and low-frequency radio emission due to coherent plasma motions in reconnecting current sheets (\S~\ref{subsec-radio}).
We then (in \S~\ref{sec-rad_reaction}) talk about the underlying physics of the radiation reaction force with the focus on a few astrophysically most relevant radiation mechanisms: synchrotron, curvature, and  inverse-Compton. 
After this, we begin a systematic exposition of the different manifestations of the effects that the radiation reaction force can have on reconnection. 
We begin this program with the discussion of a quintessential kinetic effect of radiation: the limits that radiation reaction places on nonthermal particle acceleration in collisionless magnetic reconnection (\S~\ref{sec-particle_acceleration}); and we specifically touch upon two particularly important astrophysical contexts where this issue plays a central role: the Crab Nebula gamma-ray flares (\S~\ref{subsec-Crab-flares}) and coronae of accreting black holes (\S~\ref{subsec-BH_ADC}). 
Then, we move on to the fluid picture and discuss several fluid-level effects. 
The first category of such effects concerns the effect of radiation reaction force on random thermal motions of the particles in the reconnection layer; it thus affects the layer's thermodynamics and can be described as radiative cooling (\S~\ref{sec-rad_cooling}); here, one has to distinguish the optically thin (\S~\ref{subsec-rad_cooling-thin}) and the optically thick regimes (\S~\ref{subsec-rad_cooling-thick} and~\ref{subsec-rad_pressure}).  
The second group of radiative effects on fluid-level description of magnetic reconnection deals with the effect of the radiation reaction force on the plasma bulk flows and thus concerns the dynamics and electrodynamics of the process; it constitutes the subject of~\S~\ref{sec-rad_drag}. The main two processes included under this rubric are radiative resistivity (\S~\ref{subsec-rad_resistivity}) and radiative braking of the reconnection outflow (\S~\ref{subsec-rad_drag-outflow}).  Finally, \S~\ref{sec-other} is devoted to a brief discussion of a few other, more exotic effects that take place in optically thick reconnection layers, such as radiation pressure, radiative viscosity, hyper-resistivity, and pair creation.  We then summarize the paper and discuss critical open questions and outline the directions for future research in \S~\ref{sec-conclusions-outlook}.

\section{Passive Radiative Signatures of Reconnection}
\label{sec-passive-rad-signs}

\subsection{General remarks}
\label{subsec-passive-general}

Although most of this review is devoted to a discussion of radiative magnetic reconnection, which, once again, is defined here as a situation where various aspects of the reconnection process are significantly affected by radiative energy losses, before we proceed with that discussion we first would like to discuss what one can call {\it passive} radiative signatures of reconnection.  Thus, in this section we address the question of how a reconnection region  {\it looks like}, in the literal sense of the word, even in the case where radiative back reaction on the reconnection process, discussed in the subsequent sections, is not important energetically or dynamically. Technically, this means that the radiative cooling time for the particles energized by the reconnection process is longer than the duration of the reconnection process itself, or at least longer than the characteristic time a typical particle spends inside the reconnection region (which could be taken as the Alfv\'en transit time along the layer).  
In contrast to the high-energy astrophysical reconnection discussed in the rest of this article, this non-radiative reconnection situation is the main subject of conventional reconnection research, aimed at traditional applications  such as solar flares, magnetospheric substorms, tokamak sawtooth crashes, and dedicated laboratory reconnection experiments.  The reasons why radiative signatures of reconnection have so far been neglected in these traditional contexts are that, first, the amount of light produced during these reconnection processes is very small and often not easily detectable with our current technology, and second, we have other, better diagnostic tools to study these reconnection processes, e.g., with direct measurements.  This is especially the case for reconnection in the Earth magnetosphere, where direct in-situ measurements of various plasma properties with spacecraft are available (see, e.g., the chapters by Cassak \& Fuselier and by Petrukovich et al. in this volume), and for dedicated laboratory experiments where one can use probe arrays (see the Chapter by Yamada et al. in this volume).

In contrast, in many astrophysical situations remote telescopic observations of radiation powered by a reconnection event provide our only diagnostic probe into the physical processes at play in a given system. 
In other words, we usually have no other tools to measure the plasma properties in remote astrophysical  reconnection regions and therefore, in order to make any sense of the observations, it is imperative to build predictive theoretical models that can connect the underlying physics of reconnection with the resulting light that we then can, and do, observe at Earth. 

It is interesting to note that this is in fact also true for solar flares, but this case is complicated because the post-reconnection conversion of the particle energy into radiation is neither immediate nor straight-forward, it involves interaction of the accelerated particle beams with the solar surface, chromospheric evaporation, etc.. All these additional processes bring in extra modeling uncertainties and thus, to some degree, complicate the task of using the observed flare radiation to learn something about the underlying reconnection process and about the plasma properties before reconnection.  In contrast, however, in many (although not all) astrophysical contexts reconnection happens in a free plasma flow far from any dense bodies, and in this case one can be reasonably certain that the observed radiation is in fact produced by the plasma that has been energized in a reconnection region. Most notable astrophysical examples of such a situation are astrophysical jets and winds, AGN radio-lobes, PWN, interstellar medium (ISM), and intra-cluster medium (ICM).   The relative slowness of the radiative cooling time compared with the expected reconnection (and hence particle acceleration) time enables one to  disentangle the particle acceleration and radiative cooling processes in this case. Then, a comparison of the observed flare spectrum (and its time evolution due to cooling) with the results of a rather straight-forward spectral modeling calculation can yield an unambiguous information about the energy spectrum of particles accelerated by the reconnection process. 

It is worthwhile to comment on what concrete radiative signatures of reconnection are regarded as being of interest.  The answer  depends on the specific astrophysical context and also on our observational capabilities. 
In general, it would be interesting to be able to calculate from first principles the actual spatially- and temporally resolved image of the reconnection layer (at different photon frequencies), that is, to produce a simulated picture or a movie of how a reconnection layer looks like.  
In many cases, however, the flaring region is spatially unresolved with our present technology and thus appears as a point-like source. At the same time, however, we often do have detailed spectral and temporal information.  In this situation, one is interested in (time-resolved) photon energy spectra and/or (energy-resolved) light-curves.  

As an illustration, in the next subsection we discuss the recent numerical results on nonthermal particle acceleration in relativistic pair-plasma reconnection and the corresponding radiative (synchrotron) spectral signatures. These expected radiation spectra produced by reconnection events can potentially be compared with real observations of various astrophysical systems, as was done, for example in a series of studies of reconnection-powered synchrotron flares in accreting black holes (both microquasars and AGN) by E.~de Gouveia dal Pino and her collaborators \citep{deGouveia_etal-2005, deGouveia_etal-2010, Khiali_etal-2015a, Khiali_etal-2015b, Kadowaki_etal-2015, Singh_etal-2015}.

\subsection{Radiative Signatures of Reconnection-Powered Nonthermal Particle Acceleration}
\label{subsec-passive-nonthermal}

In many high-energy astrophysical sources the observed radiation spectra and thus the energy distributions of the radiating particles are often non-thermal, described by power laws. For this reason, the question of nonthermal particle acceleration is a key problem in modern plasma astrophysics and reconnection is widely seen as one of key candidate mechanisms (see, e.g., \citep{Hoshino_Lyubarsky-2012} for review).  Particle acceleration in reconnection is usually addressed by means of PIC simulations, together with analytical theory. 

A substantial number of non-radiative PIC studies have attacked this problem in the context of relativistic reconnection in pair plasmas \citep{Zenitani_Hoshino-2001, Zenitani_Hoshino-2005, Zenitani_Hoshino-2007, Zenitani_Hoshino-2008, Jaroschek_etal-2004a, Lyubarsky_Liverts-2008, Liu_etal-2011, Bessho_Bhattacharjee-2012, Kagan_etal-2013, Cerutti_etal-2012b, Sironi_Spitkovsky-2011, Sironi_Spitkovsky-2014, Guo_etal-2014, Werner_etal-2014, Nalewajko_etal-2015}.  
In addition, \citep{Werner_etal-2013}  and \citep{Melzani_etal-2014b} have explored particle acceleration by relativistic reconnection in electron-ion plasmas, and \citep{Jaroschek_Hoshino-2009} and \citep{Cerutti_etal-2013, Cerutti_etal-2014a, Cerutti_etal-2014b} have investigated the case of relativistic pair plasma reconnection with synchrotron radiation reaction. 

It is very encouraging to see that continuing improvement in the available computer power has allowed researchers to tackle this problem in a reliable and systematic way with the adequate and necessary dynamic range  (e.g., \citep{Sironi_Spitkovsky-2014, Guo_etal-2014, Melzani_etal-2014b, Werner_etal-2014}).  
These studies have shown that relativistic reconnection does indeed efficiently generate hard power-law particle distributions, $f(\gamma)  \sim \gamma^{-\alpha}$.  The power-law index~$\alpha$ in general varies with the system size $L$ and the upstream magnetization $\sigma\equiv B_0^2/4 \pi h$, where $h$ is the relativistic enthalpy density.  For large enough pair-plasma systems, $\alpha(L,\sigma)$ seems to approach an asymptotic value $\alpha_*(\sigma)$ in the limit $L\rightarrow \infty$.  Viewed as a function of~$\sigma$, this value can be larger than $\sim$2 for modestly relativistic cases ($\sigma \sim 1$) but then monotonically decreases with $\sigma$ and asymptotically approaches a finite value $\alpha_* \simeq 1-1.2$ in the ultra-relativistic limit $\sigma \gg 1$  (e.g., \citep{Sironi_Spitkovsky-2014, Guo_etal-2014, Werner_etal-2014}).  This value is consistent with analytical predictions by \citep{Larrabee_etal-2003} and with the results of several previous numerical studies, e.g., \citep{Jaroschek_etal-2004a, Lyubarsky_Liverts-2008}. 
One of the most important new results is that relativistic non-thermal particle spectra seem to be produced at late times with high efficiency in both 2D and 3D simulations and both with and without a guide magnetic field \citep{Sironi_Spitkovsky-2014, Guo_etal-2014}. This is in contradiction with the previous picture proposed by Zenitani \& Hoshino \citep{Zenitani_Hoshino-2007, Zenitani_Hoshino-2008}, who suggested that a strong guide field is essential for nonthermal particle acceleration in 3D because it can suppress the relativistic drift-kink instability that leads to strong magnetic dissipation but inhibits nonthermal particle acceleration.   

Furthermore, as was shown by \citep{Werner_etal-2014} in 2D PIC simulations without a guide field starting with a relatively cold initial background plasma, the resulting final nonthermal power law is truncated at high energies by a combination of two cutoffs: 
\beq
f(\gamma) \sim \gamma^{-\alpha} \, e^{-\gamma/\gamma_{c1}} 
e^{-(\gamma/\gamma_{c2})^2}, 
\eeq
where the exponential cutoff $\gamma_{c1}$ and the super-exponential cutoff $\gamma_{c2}$ are well fit as functions of $L$ and~$\sigma$ by
\beq 
\gamma_{c1}  \simeq   4 \,  \sigma \, ,  \qquad 
\gamma_{c2}  \simeq  0.1 \, L/\rho_0 \, .
\eeq
Here $\rho_0 \equiv m_e c^2 /e B_0$ is the nominal Larmor radius and $\sigma \equiv B_0^2/4 \pi n_b m_e c^2$ is the "cold" background plasma magnetization, with $B_0$ being the reconnecting magnetic field strength and $n_b$ being the total (electrons and positrons) background plasma density, both taken upstream of the layer.  Also note that the length-scale $L$ in the above expression is the size of the computational domain (with aspect ratio $L_x/L_y = 1$) and is about twice the actual length of the reconnection layer.  

The first of the above two cutoffs can be understood as arising from the typical acceleration time $\ell/c$ that a given particle spends in an elementary marginally stable current layer of width $\delta \sim \rho(\bar{\gamma}) = \bar{\gamma} \rho_0 \sim \sigma \rho_0$, where $\bar{\gamma} \sim 0.3 \sigma$ is the average dissipated energy per particle, and of length $\ell \sim 50-100\, \delta$ dictated by the stability condition for secondary tearing.  The second cutoff probably arises from the finite time the particle spends in the entire layer of system-size length~$L$.  In practice, it is the smaller of the two cutoffs that matters for limiting the extent of the power law \citep{Werner_etal-2014}.  Their ratio can be expressed as
\beq
{\gamma_{c2}\over {\gamma_{c1}}}  \simeq {1\over 40} \, {L\over{\sigma\rho_0}} = 
{1\over 40} \, \cdot {{3 B_{\rm cl}}\over{2 B_0} }\, \tau_T \, ,
\eeq
where $\tau_T \equiv n_b L \sigma_T$ is the Thomson optical depth along the layer [here, $\sigma_T = (8\pi/3)\, r_e^2$ is the Thomson cross section, $r_e = e^2/m_e c^2 \simeq 2.8 \times 10^{-13}\, {\rm cm}$ is the classical electron radius], and $B_{\rm cl} \equiv e/r_e^2 = m_e^2 c^4/e^3 \simeq 6 \times 10^{15} \, {\rm G}$ is the critical classical magnetic field strength. 

The fact that there are two cutoffs allows one to define and distinguish between two different physical regimes: 
(1) the small-system regime ($L/\sigma \rho_0 \lesssim 40$), in which $\gamma_{c2} <  \gamma_{c1}$ and so $\gamma_{c2} \propto L$ determines where the power law ends, 
and 
(2) the plasmoid-dominated, large-system regime ($L/\sigma \rho_0 \gtrsim 40$), in which $\gamma_{c1} <  \gamma_{c2}$ and so $\gamma_{c1} \propto \sigma$ limits the high-energy extent of the power law, independent of~$L$.  

Next, to the extent that we are interested in potentially observable radiation signatures of reconnection, it is interesting to ask what radiation spectra are emitted by the particle distributions described above. 
If relativistic reconnection indeed produces a power law energy spectrum of electrons with an index $\alpha \simeq 1.2$ in the ultra-relativistic, high-$\sigma$ regime, then the corresponding synchrotron photon spectrum {\it immediately after the reconnection event} will be a nearly flat power law with a spectral index of $(\alpha-1)/2 \sim 0.1$, which in practice would be indistinguishable from a flat spectrum. In terms of the photon-number power-law index $\Gamma_{\rm ph}$, defined by $n_{\rm ph} (\epsilon_{\rm ph}) \sim \epsilon_{\rm ph}^{-\Gamma_{\rm ph}}$, this corresponds to $\Gamma_{\rm ph} = (\alpha+1)/2 \simeq 1.1$. 
In large systems ($L/\sigma \rho_0 \gtrsim 40$), the power-law synchrotron spectrum is then expected to extend up to the characteristic photon energy of 
\begin{eqnarray}
\epsilon_{\rm sync,\, max} &=&  {3\over 2} \, \hbar \Omega_{c0}\, \gamma_{c1}^2 = 
{3\over 2} \, \hbar {e B_0\over{m_e c}}\, \gamma_{c1}^2  = 
{3\over 2} \, \alpha_{\rm fs}^{-1}\, m_e c^2\, {r_e\over{\rho_0}} \, \gamma_{c1}^2 \nonumber \\
& = & {3\over 2} \, \alpha_{\rm fs}^{-1}\, m_e c^2\, {B_0\over{B_{\rm cl}}} \, \gamma_{c1}^2 \simeq
100\, {\rm MeV} \, {B_0\over{B_{\rm cl}}} \, \gamma_{c1}^2 \, , 
\end{eqnarray}
where $\alpha_{\rm fs} = e^2/\hbar c \simeq 1/137$ is the fine structure constant. 
Substituting our expression $\gamma_{c1} \simeq 4 \sigma$, we find 
\beq
\epsilon_{\rm sync,\, max} \simeq  
24\, \alpha_{\rm fs}^{-1}\, m_e c^2\, {B_0\over{B_{\rm cl}}} \,  \sigma^2   \, , 
\eeq
It is interesting to note that this limit grows very rapidly with the magnetic field, namely as $B_0^5$. 

On a longer time scale following a reconnection event, subsequent cooling evolution will, of course, soften the emission spectrum since the highest energy particles have shorter radiation cooling times: 
\begin{eqnarray}
t_{\rm cool}^{\rm sync} &=& {{\gamma m_e c^2}\over{P_{\rm rad}}} = 
{{\gamma m_e c^2}\over{(4/3)\, \sigma_T c B^2/8\pi}} = 
{9\over 4}\, (\gamma\, \Omega_{c0})^{-1} \, {\rho_0\over{r_e}} \nonumber \\
&= & {9\over 4}\, (\gamma\, \Omega_{c0})^{-1} \, {B_{\rm cl}\over{B}}  \simeq 
7.7 \times 10^8 \, {\rm s}\, \gamma^{-1} \, \biggl( {B\over{1\, {\rm G}}} \biggr)^{-2} \simeq 
24\, {\rm yr}\, \gamma^{-1} \, \biggl( {B\over{1\, {\rm G}}} \biggr)^{-2} \, .
\end{eqnarray}
Here, $\Omega_{c0} \equiv e B/m_e c \simeq 1.76 \times 10^7 \, (B/{1\,{\rm G}})\, {\rm rad/s}$ is the classical electron cyclotron frequency. 
This results in a time-evolving cooling energy limit $\gamma_{\rm br}$ at the particle energy set by 
$t = t_{\rm cool}(\gamma_{\rm br})$, above which the particle energy spectrum is cut off sharply. 

Next, in the case of a complex system (e.g., a corona) with a large number of independent reconnection events (flares) continuously injecting power-law populations of energetic relativistic electrons, each with an initial power-law index of $\alpha_{\rm inj} = 1.2$, one expects the interplay of this continuous injection and synchrotron radiative cooling to result in a steady-state electron distribution with a power-law index 
$\alpha_{\rm ss} = 1+\alpha_{\rm inj} \simeq 2.2$.  This corresponds to a photon number index of $\Gamma_{\rm ph} = (1+\alpha_{\rm ss})/2 =1.6$.  Similar considerations apply in the case of IC emission, resulting in the same photon index of $\Gamma_{\rm ph} =1.6$ for the IC photons, which is intriguingly close to the hard X-ray photon index of 1.7 often observed in the low-hard state of~XRBs (e.g., \citep{Remillard_McClintock-2006}).

Finally, when thinking about possible observable radiative signatures of relativistic reconnection at highest photon energies (hence produced by the highest-energy accelerated particles), one should take into account a possible anisotropy of the accelerated particle population.  As was shown by \citep{Cerutti_etal-2012b}, the highest energy particles accelerated in a reconnection layer may be focused in a few tight beams that sweep from side to side, while staying mostly in the current sheet plane. This {\it kinetic beaming} effect is strongly energy-dependent, with the effective solid angle $\Omega$ of the particle population decreasing from $\Omega/4\pi \sim 1$ for low- and modest-energy particles to as small as $\Omega/4\pi \sim 10^{-2}$ for the highest-energy ones.  
This effect potentially has important implications for understanding radiative signatures of reconnection and, especially, for connecting theoretical models with observations since it suggests that the usual isotropic emission assumption may lead to large errors in evaluating the energetic requirements implied by the observed radiation flux.
Kinetic beaming is also important for correctly interpreting the very rapid emission variability frequently observed in many relativistic astrophysical sources, such as the Crab PWN and blazar and GRB jets. This is because the radiative flux as seen by an external observer is greatly enhanced (relative to the isotropically-averaged total flux) when one of the beams intersects the observer's line of sight. As a result, the observed signal is strongly intermittent, leading to an enhanced rapid and energy-dependent variability.


\subsection{Loop-top hard X-ray emission in solar flares}
\label{subsec-loop_top}

An important example of high-energy radiation produced promptly by the plasma energized in a reconnection event  is the hard X-ray (up to about 100 keV) emission at the top of post-reconnected coronal magnetic loops in solar flares. In contrast to the hard X-ray emission produced at the footpoints of the reconnected loops on the solar surface, which is traditionally understood as bremsstrahlung radiation emitted by the electrons accelerated in the coronal reconnection region as they  strike a dense cold target (the solar photosphere), the loop-top emission involves only those plasma particles that have gone through, and have been accelerated in, the reconnection region, without the agency of any other plasma.  This radiation is also believed to be optically-thin nonthermal bremsstrahlung corresponding to a power-law distribution of electrons extending up to relativistic (MeV) energies (Krucker et al. 2010), although it can also be modeled as a kappa-distribution \citep{Oka_etal-2013, Oka_etal-2015}. 
The observed X-ray radiative power is so high that it implies an extremely high efficiency of non-thermal particle acceleration, with the number density of energetic particles populating the power-law tail being comparable to the expected density of ambient thermal particles. This challenges traditional flare emission models and strongly suggests that a large fraction of the ambient plasma particles in the flare region are accelerated into the power-law tail (e.g., \citep{Krucker_etal-2010, Oka_etal-2013, Oka_etal-2015}).  However, these challenges may be partially alleviated by noticing that the plasmoid-dominated reconnection regime expected in solar flares naturally leads to strong inhomogeneity of the energized plasma, concentrating it into relatively compact, dense plasmoid cores.  Since bremsstrahlung is a collisional process, with radiated power per unit volume proportional to the square of the plasma density, this inhomogeneity can greatly enhance the overall emission power.  This effect can be easily tested in PIC simulations and perhaps also in laboratory laser-plasma studies of reconnection.

It is interesting to try to generalize Werner et al's (2014) results for the high-energy nonthermal cutoff of particles accelerated by a relativistic pair-plasma reconnection process described in the preceding section to the case of non-relativistic reconnection in electron ion plasmas and to apply them to solar flares. 
Since the flaring region size  in solar flares ($10^9-10^{10}$~cm) is many orders of magnitude larger than the ion Larmor radius, one is squarely in the large-system regime. Therefore, one expects that the relevant cutoff  is $\epsilon_{c1}$, set by the acceleration in elementary (marginally stable to secondary tearing) current layers, 
$\epsilon_{c1} \simeq e E_{\rm rec} \ell$. 
We can estimate the characteristic length of these elementary layers as 
$\ell \sim 100\, \delta  \sim 100\, \rho_{i,\rm layer}$, where the layer thickness $\delta$ is taken to be comparable to the Larmor radius of the ions inside the layer~$\rho_{i,\rm layer}$.  
Taking for illustration $B_0 = 100\, {\rm G}$ and the plasma density in the layer $n_e = 10^{10}\, {\rm cm^{-3}}$, and hence $V_A \simeq 2\times 10^3 \, {\rm km/s} \simeq 0.7 \times 10^{-3}\, c$, one can estimate (e.g., from the pressure balance across the layer) the plasma temperature in the layer as 
$k_B T = B_0^2/(16 \pi n_e)  \simeq 12\, {\rm keV}$. This corresponds to an ion Larmor radius, and hence the elementary layer's thickness, of $\delta\sim \rho_{i,\rm layer} \simeq 1\, {\rm m}$, and hence the elementary layer length of $\ell \sim 100\, {\rm m}$. 
Next, since we are dealing with non-relativistic collisionless reconnection, the reconnection electric field can be estimated as $E_{\rm rec} \simeq 0.1\, B_0 V_A/c \simeq 0.07\, {\rm G}$. Therefore, the expected high-energy cutoff should be $\epsilon_{c1} \simeq e E_{\rm rec} \ell \sim 200\, {\rm kev}$, corresponding to mildly relativistic electrons.


\subsection{Coherent Radio Emission}
\label{subsec-radio}

In addition to the production of high-energy radiation through incoherent mechanisms such as synchrotron, IC, and bremsstrahlung radiation, another important radiative aspect of reconnection is the possible generation of coherent low-frequency (e.g., radio or microwave) emission associated with collective plasma motions.  This emission may be driven by various small-scale plasma motions excited inside thin reconnection current layers by various plasma  instabilities, such as the secondary tearing, the drift-kink, lower-hybrid instability, ion-acoustic, and/or Buneman  instabilities, etc..  The nonlinear development of these instabilities may lead to the production of a broad spectrum of fluctuations, e.g., having the form of plasmoids and flux ropes of different sizes in the case of the secondary tearing instability. These plasmoids exhibit complex dynamics marked by their interaction with each other through mergers.  
It is then plausible that some of these fluctuations will eventually be converted into low-frequency electromagnetic waves that may escape the system and be observed at Earth. 
Although the efficiency of conversion of reconnection-released energy into such low-frequency emission can be  low, this emission may nevertheless provide an important additional diagnostics window into the reconnection process. 
The typical frequencies of this radiation are expected to be low, on the scale of a fraction of the plasma frequency and lower, corresponding to radio emission in systems as diverse as the solar corona \citep{Shklovsky-1947} and the Crab pulsar magnetosphere~\citep{Uzdensky_Spitkovsky-2014}.  In coronae of accreting black holes in XRBs, however, the plasma density (and hence the plasma frequency) is much higher and hence the corresponding emission probably falls into the infra-red or even optical range (\citep{Goodman_Uzdensky-2008}).  

\section{Radiation Reaction Force}
\label{sec-rad_reaction}

The main reason why radiation can sometimes be important in various plasma processes, including reconnection, is that it affects the motion of the plasma particles and thus influences the basic dynamics and energetics of the process in question. The primary effect of radiation can be described by the radiation-reaction drag force ${\bf f}_{\rm rad}$ experienced by the individual single particles and, associated with this force, the energy loss term $P_{\rm rad}$ in the particle's energy equation.  

The relativistic 4-force representing radiation reaction on an emitting particle, called the Abraham-Lorentz-Dirac (ALD) force, can be written as (\citep{Jackson-1975}):
\beq
F_{\rm rad}^\mu = {2e^2\over{3c^3}} \, {{d^2 u^\mu}\over{d\tau^2}} - {P_{\rm rad}\over{c^2}}\, u^\mu \, .
\label{eq-rad-4-force}
\eeq
where $u^\mu$ is the particle's 4-velocity, $\tau = \gamma^{-1} t$ is the particle's proper time, and where the radiative power $P_{\rm rad}$ is given by the Larmor formula, which reads, in relativistically covariant notation \citep{Jackson-1975}:
\beq
P_{\rm rad} = {2\over 3} \, {e^2 \over{m^2 c^3}} \, {{d p_\mu}\over{d\tau}} {{d p^\mu}\over{d\tau}} \, . 
\eeq
Here, $p^\mu = m u^\mu = m c (\gamma, \gamma \pmb{\beta})$ is the four-momentum of the particle moving with a 3-velocity ${\bf v} = \pmb{\beta} c$. 
This expression can be recast in terms of the parallel ($a_\parallel$) and perpendicular ($a_\perp$) (with respect to the particle direction of motion) 3-acceleration as (\citep{RL-1979}): 
\beq
P_{\rm rad} = {2\over 3} \, {e^2 \over{c^3}} \gamma^4\, (a_\perp^2 + \gamma^2 a_\parallel^2) \, . 
\eeq
Furthermore, for a particle moving non-relativistically, this formula reduces to the familiar non-relativistic Larmor formula: 
\beq
P_{\rm rad} = {2\over 3} \, {e^2 \over{m^2 c^3}}\, |\dot{\bf p}|^2 = 
{2\over 3} \, {e^2 a^2\over{c^3}}  \, , 
\eeq
where $a$ is the particle's 3-acceleration. 

Returning to our discussion of radiation reaction, in the more familiar 3D language the radiation reaction enters the relativistic equation of motion of a charged particle as an additional friction force~${\bf f}_{\rm rad}$: 
\beq
d{\bf p}/{dt} = q \, ({\bf E} + [{\bf v\times B}]/c) + {\bf f}_{\rm rad} \, , 
\eeq
where ${\bf p} = m \gamma {\bf v}$ is the particle's relativistic 3-momentum. 
The radiation reaction 3-force is related to the ALD 4-force via 
$F_{\rm rad}^\mu = (\gamma \pmb{\beta} \cdot {\bf f}_{\rm rad}, \gamma {\bf f}_{\rm rad})$. 

The first term in expression~(\ref{eq-rad-4-force}) for $F_{\rm rad}^\mu$, called the Schott term, is quite peculiar: 
it involves the second time derivative of the 4-velocity and hence the third-order time derivative of position, which means that the equation of motion that includes this term becomes a third-order differential equation in time.  
This term dominates in the non-relativistic case ($|{\bf v}| \ll c$, $\gamma \rightarrow 1$), in which the radiation reaction force reduces to what is known as the Abraham-Lorentz force (e.g., \citep{Landau_Lifshitz-1971}): 
\beq
{\bf f}_{\rm rad} (\gamma \ll 1) \approx {2e^2\over{3c^3}}\, {{d^2 {\bf v}}\over{dt^2}} \, .
\eeq

In contrast, in the case of ultra-relativistic motion (of main interest to this review) the Schott term can be shown to be small.  Ignoring it, the radiation reaction 3-force on ultra-relativistic particles can be expressed in terms of the radiative power simply as
\beq
{\bf f}_{\rm rad} (\gamma \gg 1) \approx  -\, {{P_{\rm rad}}\over{c}} \, \pmb{\beta} \, . 
\label{eq-f_rad-general}
\eeq
That is, the radiation reaction force in this case indeed plays a role of a friction force, directed opposite to the particle direction of motion.  Furthermore, the magnitude of the radiation reaction force for an ultra-relativistic particle is then simply $|{\bf f}_{\rm rad}| \approx P_{\rm rad}/c$. This is consistent with the notion that the rate of work done by the radiation reaction force, ${\bf f}_{\rm rad} \cdot {\bf v} = - P_{\rm rad}\, v^2/c^2$, becomes equal to the particle's radiative energy loss rate $- P_{\rm rad}$ in the limit $v\rightarrow c$. 


We shall now apply these general expressions to several specific astrophysically-important radiative processes corresponding to different types of accelerated particle motion that enters the above formulae for~$P_{\rm rad}$. 
In astrophysical plasmas acceleration is usually due to the particle motion in an external electromagnetic field. 
The most important types of accelerated motion, and the corresponding radiation mechanisms, are: 
(1) cyclotron gyro-motion in a magnetic field and, correspondingly, the cyclotron/synchrotron emission; 
(2) parallel motion along a curved magnetic field line and curvature emission; 
(3) oscillatory motion of a charged particle in the electromagnetic field of an incident electromagnetic wave, resulting in Compton scattering (usually referred to as inverse-Compton (IC) scattering if the energy of incident photons is smaller than that of the scattering particles); in astrophysical studies focussed on production of high-energy radiation by energetic electrons scattering soft seed photons, one sometimes treats IC scattering effectively as an emission process;   
(4) motion of one charged particle in the electric field of another in a close binary collision and, correspondingly, bremsstrahlung (free-free) radiation emission. 

We shall now discuss the radiative power and the radiation reaction force for each one of these mechanisms (except for bremsstrahlung) in more detail. 


{\bf (1) Synchrotron Radiation.} 
First, for the cyclotron motion of a particle with a 4-velocity $(\gamma, \pmb{\beta} \gamma)$ in a general electro-magnetic field, the radiative power is~(\citep{Landau_Lifshitz-1971})
\beq
P_{\rm rad} \simeq {1\over{4\pi}}\, \sigma_T c \gamma^2 \, 
\biggl( ({\bf E} + [\pmb{\beta} \times {\bf B}])^2 - (\pmb{\beta} \cdot {\bf E})^2 \biggr) \, .
\eeq
This expression is actually only approximate: it is based on a perturbative approach, keeping only the acceleration due to the usual Lorentz 4-force $-(e/c) u_\nu F^{\mu\nu}$ in the Larmor formula, while neglecting the effect of the radiation reaction force itself. However, it is valid in most realistic astrophysical situations.

In the frame of reference in which the electric field vanishes (the so-called Teller-Hoffmann frame), the radiative power of a charged particle spiraling in a magnetic field is
\beq
P_{\rm rad} = P_{\rm synch} = 2 \sigma_T c \,  \beta^2 \gamma^2 \,{{B_\perp^2}\over{8\pi}} = 
{2\over 3} r_e^2 c \beta^2 \gamma^2 \,{B^2} \, \sin^2 \alpha  \, , 
\eeq
where $\alpha$ is the pitch angle of the particle relative to the direction of the magnetic field and 
$B_\perp \equiv B \sin \alpha$ is the magnetic field component perpendicular to the particle's velocity. 
This radiation is called synchrotron radiation in the case of ultra-relativistic particles, cyclotron radiation in the case of non-relativistic particles, and gyro-synchrotron radiation for the intermediate case of moderately relativistic particles. 

It is important to note that cyclotron gyration is perpendicular to the magnetic field and hence only the perpendicular velocity of the particle is involved in this radiation. Even a very energetic particle in a strong magnetic field produces no synchrotron radiation if it moves strictly parallel to the field (although it may still produce the so-called curvature radiation if the field lines are curved, see below). 

In the ultra-relativistic case $\gamma \gg 1$, the synchrotron radiative power (in the Teller-Hoffmann frame) becomes 
\beq
P_{\rm synch}(\gamma\gg 1) = 2 \sigma_T c  \gamma^2 \,{B^2\over{8\pi}}\, \sin^2 \alpha \, , 
\eeq
and hence the corresponding synchrotron radiation reaction force is 
\beq
{\bf f}_{\rm rad}^{\rm synch} = - {{\bf v}\over c^2} P_{\rm synch}   = 
- 2 \, \pmb{\beta}\, \sigma_T \, \gamma^2 \,{B^2\over{8\pi}}\, \sin^2 \alpha \, .
\eeq 

It is worth noting that $P_{\rm synch}$ is proportional to the square of the particle energy and hence the radiative cooling time, 
$t_{\rm cool} = \gamma m_e c^2/P_{\rm synch} \sim \gamma^{-1}$, 
is inversely proportional to the particle energy. 
Correspondingly, ${\bf f}_{\rm rad}^{\rm synch}$ and synchrotron energy losses are is especially important for highest-energy relativistic particles. 

It is also important to note that the above simple expressions for $P_{\rm synch}$ and ${\bf f}_{\rm rad}^{\rm synch}$ are valid only in the Teller-Hoffmann frame, where electric field vanishes. This reference frame corresponds to the ${\bf E} \times {\bf B}$ drift, ${\bf v}_E = c\, [{\bf E\times B}]/B^2$.  An important consequence is that synchrotron emission arises only due to the perpendicular (to ${\bf B}$) motion of particles relative to the ${\bf E\times B}$ drift.  In particular, this means that a cold ideal-MHD plasma flow with the ${\bf E\times B}$ velocity does not produce synchrotron emission,  
even if it is highly relativistic as in the case of a pulsar wind. 

Also, the Teller-Hoffmann frame exists only if the electric field is weaker than the magnetic field and has no component parallel to~${\bf B}$, as can be seen by examining two electromagnetic-field Lorentz invariants, ${\bf E\cdot B}$ and $E^2-B^2$. If these conditions are not satisfied, then the radiative power and hence the radiation reaction force need to be found from more general expressions for the ALD force. 

We finally note that the above formula for synchrotron radiation is valid only if the magnetic field remains smooth on the length scale of radiation formation, which is about $\rho_L /\gamma = \rho_0$. 
This requirement is usually satisfied in most astrophysical cases, but there are situations where it is violated, namely,  when the nominal Larmor radius $\rho_0$ is larger than the magnetic field reversal scale~$\lambda_B$.  
In this case, one has the so-called jitter radiation \citep{Medvedev-2000} instead of synchrotron, which may be relevant for GRB prompt emission.
However, although the jitter radiation spectrum differs substantially from that of the classical synchrotron radiation, it turns out that the overall radiative power and hence the radiation reaction force are the same in the two cases. 


{\bf (2) Curvature radiation.}
In some astrophysical applications, especially involving ultra-relativistic particles moving in a strong magnetic field, e.g., in a pulsar magnetosphere, the particles quickly lose their perpendicular (cyclotron) energy by synchrotron radiation and fall into their lowest Landau level.  Then their subsequent motion becomes essentially one-dimensional (1D), parallel to the magnetic field, and is adequately described as the motion of bids on a wire or train cars running along the rails (\citep{Sturrock-1971}). 
However, this does not mean that the motion becomes completely trivial or that radiative effects are not important. In particular, if the magnetic field lines are not straight, the particles still experience centripetal acceleration as they move along the curved field lines, and hence can still radiate according to the Larmor formula (\citep{Shklovsky-1960}). For relativistic motion along curved magnetic fields lines this radiation is called the curvature radiation and its radiative energy loss rate is given by (\citep{Shklovsky-1960, Sturrock-1971, Chugunov_etal-1975, Zheleznyakov-1977}): 
\beq
P_{\rm rad} = P_{\rm curv} = {2\over 3} \, {c e^2\over{R_c^2}}\, \gamma^4 \, , 
\eeq
where $R_c$ is the field lines' radius of curvature. 

Correspondingly, ultra-relativistic particles experience a radiative reaction drag force: 
\beq
{\bf f}_{\rm rad}^{\rm curv} = -\, {2\over 3} \, {e^2\over{R_c^2}}\, \gamma^4 \, \pmb{\beta}.
\eeq


{\bf (3) Inverse-Compton radiation.}\\
In the case of Compton scattering in an isotropic radiation field, the radiative power $P_{\rm rad}$ entering the above expressions (\ref{eq-rad-4-force}) and (\ref{eq-f_rad-general}) for the radiation reaction force for relativistic electrons is given by (e.g., \citep{Blumenthal_Gould-1970, Pozdnyakov_etal-1983}):
\beq
P_{\rm rad} =  P_{\rm IC}  = (4/3)\, \sigma  c  \, U_{\rm rad} \, \gamma^2 \beta^2 
\approx (4/3)\, \sigma  c  \, U_{\rm rad} \, \gamma^2 \, , 
\label{eq-Prad-IC}
\eeq 
corresponding to the radiation reaction force
\beq
{\bf f}_{\rm rad}^{\rm IC}  = - {\bf v}\, P_{\rm rad}/c^2 
= - \,(4/3)\, \sigma  \gamma^2 \, U_{\rm rad} \, {\bf v}/c \, ,
\label{eq-f_rad-IC}
\eeq
where $U_{\rm rad}$ is the radiation energy density and $\sigma$ is the applicable scattering cross-section.

In most astrophysical applications the scattering is in the so-called Thomson regime, in which the seed photon's energy in the electron's rest frame, $\epsilon_{\rm ph}' \sim \gamma \epsilon_{\rm ph,\ seed}$, is less than $m_e c^2$; then one can use the simple energy-independent Thomson cross-section, $\sigma = \sigma_T$. 
In the opposite case, however, one has to use a more general quantum-mechanical Klein-Nishina expression for the cross-section: 
\beq
\sigma_{\rm KN}(x) = {3\over 4}\, \sigma_T\, 
\biggl[ {{1+x}\over{x^2}} \, \biggl( {{2(1+x)}\over{1+2x}} - {{\ln(1+2x)}\over{x}} \biggr)
+ {{\ln(1+2x)}\over{2x}} - {{1+3x}\over{(1+2x)^2}} \biggr] \, ,
\eeq
where $x \equiv \epsilon_{\rm ph}' / m_e c^2$. 
In the ultra-relativistic limit $x \gg 1$, this expression can be approximated as
\beq
\sigma_{\rm KN}(x \gg 1) \approx {3\over 8}\, \sigma_T\, {{\ln 2x + 1/2}\over x} \, .
\eeq

Since the radiative reaction force on relativistic particles due to both synchrotron and inverse-Compton (in the Thomson regime) mechanisms scales as the square of the particle energy $\epsilon=\gamma m c^2$, the corresponding radiative cooling time, $\tau_{\rm rad} = \epsilon/P_{\rm rad}$ scales inversely with the energy. 
For curvature radiation the effect is even stronger since $P_{\rm rad}^{\rm curv} \sim \gamma^4$ and hence $\tau_{\rm rad}^{\rm curv} \sim \gamma^{-3}$.  This means that for each of these processes radiative losses affect more energetic particles the most. 
What this implies is that radiative energy losses lead not only to the overall cooling of the plasma but also affect the shape of the particle distribution function.  In particular, radiative losses may result in an effective upper energy limit on nonthermal particle acceleration, which may have very important observational consequences, as we discuss in the next section. 

\section{Radiation Effects on the Kinetic Picture of Magnetic Reconnection: 
Limiting Nonthermal Particle Acceleration}
\label{sec-particle_acceleration}

Nonthermal particle acceleration, the hallmark of which is usually considered to be the production of (truncated) power-law particle energy distributions, is an important and ubiquitous phenomenon in collisionless space-, solar-, and astrophysical plasmas.  Among plasma-physical processes commonly believed to be responsible for nonthermal particle acceleration, the most popular are collisionless shocks, 
magnetic reconnection, and MHD turbulence.  Whatever the mechanism is, however, a particle's energy can be increased only by the work done by the electric field, since the Lorentz force due to the magnetic field is perpendicular to the particle's direction of motion.  It therefore follows that, in the presence of radiative losses, the maximum Lorentz factor $\gamma_{\rm rad}$ that a charged particle accelerated by an electric field $E$ can attain is determined by the balance between the accelerating electric force $eE$ and the radiation reaction force $f_{\rm rad}(\gamma)$, i.e., $f_{\rm rad}(\gamma_{\rm rad})= eE$.  Importantly, the electric field in most astrophysical applications is tied to the magnetic field and is typically of order the motional electric field, $E\lesssim v B/c = \beta B$, where $v\equiv \beta c$ is the plasma 3-velocity.  Thus, it is often useful to parametrize the electric field in terms of the magnetic field, e.g., by introducing a dimensionless parameter $\beta_E = E/B$, which is usually less than unity. 
In most magnetically dominated systems, the typical flow velocity is of the order of the Alfv\'en speed, $V_A$, and so one typically expects $\beta_E \lesssim V_A/c$. 
For example, in the context of magnetic reconnection, the relevant electric field is usually the main reconnection electric field and $B$ is the reconnecting component of the magnetic field; then, $\beta_E = \beta_{\rm rec} = v_{\rm rec}/c$, where the $v_{\rm rec}$ is the reconnection inflow velocity, typically of order $v_{\rm rec} \sim 0.1 V_A$ for collisionless reconnection (e.g., \citep{Birn_etal-2001}) and $v_{\rm rec} \sim 0.01 V_A$  for resistive-MHD reconnection in the large-system, plasmoid-dominated regime \citep{Bhattacharjee_etal-2009, Huang_Bhattacharjee-2010, Uzdensky_etal-2010, Loureiro_etal-2012}, although it can be higher in the presence of background turbulence (e.g., \citep{Lazarian_Vishniac-1999, Kowal_etal-2009, Loureiro_etal-2009, Eyink_etal-2011}).

In relativistic plasmas, where $\beta_E \sim 1$, one obtains the following upper limits on relativistic particle acceleration in the presence of the three main radiative mechanisms (synchrotron, IC, and curvature) discussed in \S~\ref{sec-rad_reaction}: 

{\bf (1) Synchrotron radiation} 
(\citep{Guilbert_etal-1983, deJager_etal-1996, Lyutikov-2010, Uzdensky_etal-2011}): 
\beq
\gamma_{\rm rad}^{\rm sync} = 
{1\over{\sin\alpha}} \, \biggl[ {{3\, \beta_E}\over{2}}\, {e\over{B r_e^2}} \biggr]^{1/2} = 
{1\over{\sin\alpha}} \, \biggl[ {{3\, \beta_E}\over{2}}\, {B_{\rm cl}\over B} \biggr]^{1/2} \, , 
\label{eq-gamma_rad-synch}
\eeq
where $\alpha$ is the pitch angle of the particle with respect to the magnetic field (here we assume that the electric field is parallel to the direction of the particle's motion). The corresponding maximum characteristic synchrotron photon energy then is 
\beq
\epsilon_{\rm ph, max}^{\rm sync} = {3\over 2}\, (\gamma_{\rm rad}^{\rm sync})^2\, \hbar \Omega_{c0} = 
{9\over 4}\, \alpha_{fs}^{-1}\, m_e c^2\, \beta_E \simeq 160\, {\rm MeV} \, \beta_E\, , 
\eeq
where $\alpha_{fs} = e^2 / \hbar c \simeq 1/137$ is the fine structure constant.


{\bf (2) Curvature radiation} 
(\citep{Sturrock-1971, Chugunov_etal-1975, Lyutikov_etal-2012b}):
\beq
\gamma_{\rm rad}^{\rm curv} = \biggl({{3 E_\parallel R_c^2}\over{2e}}\biggr)^{1/4} \, , 
\label{eq-gamma_rad-curv}
\eeq
where $E_\parallel$ is the accelerating parallel (to the magnetic field and to the particle velocity) electric field. 
This corresponds to a characteristic maximum photon energy that can be achieved by curvature radiation (\citep{Lyutikov_etal-2012b}) of 
\beq
\epsilon_{\rm ph, max}^{\rm curv} 
= \biggl({3\over 2} \biggl)^{7/4}\, \hbar c R_c^{1/2} \, \biggl({E_\parallel\over e} \biggl)^{3/4} \, ,
\eeq
which can be recast as 
\beq
\epsilon_{\rm ph, max}^{\rm curv}
= \biggl({3\over 2} \biggl)^{7/4}\, m_e c^2\, \alpha_{fs}^{-1} \, \sqrt{R_c\over{r_e}}\, 
\biggl({E_\parallel\over{B_{\rm cl}}} \biggl)^{3/4} \, .
\eeq


{\bf (3) Inverse Compton radiation} 
(in the Thomson regime): 
\beq
\gamma_{\rm rad}^{\rm IC} =  
\biggl[ {3\over{4}}\, {{e E}\over{\sigma_T U_{\rm rad}}} \biggr]^{1/2} = 
\biggl[ {{9\,\beta_E}\over{32 \pi}}\, {{B B_{\rm cl}}\over{U_{\rm rad}}} \biggr]^{1/2} \, . 
\label{eq-gamma_rad-IC}
\eeq


These radiation-reaction upper energy limits become important in situations where they are lower than the applicable energy limits that may arise due to other reasons. One such other limit, for example, is due to a finite maximum available voltage drop associated with a given system size~$L$ and~electric field~$E$: 
$\gamma_{\rm max} = \epsilon_{\rm max}/m c^2 = eE L/m c^2 =  \beta_E\, L/\rho_0$, 
where, once again, $\rho_0 \equiv m_e c^2/ eB$ is the fiducial Larmor radius of a mildly relativistic electron corresponding to a magnetic field~$B$ \citep{Hillas-1984, Aharonian_etal-2002}. 
While the condition $\gamma_{\rm rad} < \gamma_{\rm max}$ is usually not satisfied in heliospheric environments, this situation does happen naturally in some of the most important high-energy astrophysical systems. In particular, this happens in pulsar magnetospheres (curvature and synchrotron radiation), in PWN (synchrotron), and in black-hole accretion flows (inverse Compton and synchrotron), as we will discuss in the following two subsections.  
For example, in the case of synchrotron radiation, the condition $\gamma_{\rm rad} < \gamma_{\rm max}$ can be recast (ignoring factors of order unity) as 
$L > \rho_0 \, (\rho_0/r_e)^{1/2} = r_e\, (B/B_{\rm cl})^{-3/2} 
\simeq 1.3 \times 10^{11}\, {\rm cm}\, [B/(1\,{\rm G})]^{-3/2} = 
1.3\, {\rm m}\, B_6^{-3/2}$, where $B_6 \equiv B/(1\,{\rm MG})$ is the magnetic field normalized to 1~MG, a value typical for gamma-ray-emitting pulsar magnetospheres near the light cylinder and for accretion disks of stellar-mass black holes in~XRBs.

An equivalent way to think about the relative importance of radiation reaction in limiting particle acceleration is to consider relativistic particles moving at the radiation reaction limit $\gamma_{\rm rad}$ and to cast their radiative cooling length, $\ell_{\rm cool} = c t_{\rm cool} \equiv \gamma m c^3 /P_{\rm rad}$ in terms of their Larmor radius, $\rho \equiv \gamma m c^2 / eB$.  
Since $\gamma_{\rm rad}$ is determined by the force balance between the radiation reaction force $f_{\rm rad} \approx P_{\rm rad}/c$, and the accelerating electric force~$eE= e \beta_E B$, one can immediately see that 
$\ell_{\rm cool} = \rho {B/E}  = \beta_E^{-1} \rho(\gamma_{\rm rad})$. 
In particular, in the case of reconnection, the electric field is parametrized in terms of the upstream reconnecting magnetic field $B_0$ as $E = \beta_{\rm rec} B_0$, where $\beta_{\rm rec}$ is the dimensionless reconnection rate which, in collisionless relativistic systems is of order~0.1. 
Thus, we see that 
\beq
\ell_{\rm cool}(\gamma_{\rm rad}) \simeq  \gamma_{\rm rad} m_e c^2/ eE \simeq \beta_{\rm rec}^{-1}\, \rho(\gamma_{\rm rad}) \, ,  
\eeq
i.e., only perhaps by a factor of $\beta_{\rm rec}^{-1} \sim 10$ longer than the Larmor radius of these particles.  This means that, if one is interested in extreme high-energy nonthermal particle acceleration, one has to take radiation reaction into account once the size of the accelerating region exceeds about 
$10\rho(\gamma_{\rm rad})$. 
Once again, this is usually not a concern in most heliospheric environments, but this situation is ubiquitous in astrophysics. 

Finally, we would like to note that the above formulation treats radiation reaction as a continuos force on the particles, ignoring the fact that in reality radiation is emitted in the form of discrete photons. When the energy of the emitted photons becomes comparable to the kinetic energy $\gamma m c^2$ of the emitting particle, one has to take the quantized, discrete nature of the radiation process into account. For example, for synchrotron radiation this happens when the particle's Lorentz  factor approaches $\gamma_Q = \rho_0/l_C = B_Q/B$, where $l_C \equiv \hbar/m c$ is the Compton length scale and $\rho_0 \equiv m c^2/e B$, and $B_Q = \alpha_{\rm fs} B_{\rm cl} \simeq 4.4 \times 10^{13} \, {\rm G}$ is the critical quantum magnetic field.  Comparing $\gamma_Q$ with $\gamma_{\rm rad}^{\rm sync}$ and neglecting for simplicity order-unity factors like $\sin\alpha$ and $3\beta_E/2$, we see that $\gamma_{\rm rad}^{\rm sync}/\gamma_Q \sim (B/\alpha_{\rm fs} B_Q)^{1/2}$. Therefore, synchrotron radiation reaction prevents a particle from reaching the quantum-radiation regime (i.e., $\gamma_{\rm rad}^{\rm sync} < \gamma_Q$) under most astrophysically-relevant circumstances, namely, as long as $B \lesssim \alpha_{\rm fs} B_Q \sim 10^{11}\, {\rm G}$.  The only class of astrophysical objects for which this inequality is violated is neutron stars and, especially, ultra-magnetized neutron stars called magnetars: typical magnetic fields in normal neutral stars are of order~$10^{12}\, {\rm G}$, and in magnetars they routinely reach $10^{15}\, {\rm G}$. 
This means that when considering energetic plasma processes, such as reconnection, in a close vicinity of these objects, the usual continuous-emission picture for synchrotron radiation is not applicable and one should instead describe it as emission of discrete quanta. 


\subsection{Radiative Relativistic Reconnection and the Crab Nebula Flares}
\label{subsec-Crab-flares}

One of the most prominent examples of possible radiative effects on nonthermal particle acceleration is given by the emission produced by the Crab PWN. 
It has now been reasonably firmly established that most of the baseline steady-state nothermal continuum emission from the Nebula, spanning from radio, to optical, to X-rays, and high-energy (tens of MeV) gamma-rays, is produced by synchrotron radiation from ultra-relativistic electrons and positrons  that populate the Nebula (e.g., \citep{Shklovsky-1957, Shklovsky-1966}). Then, however, the spectrum is observed to drop rather sharply above about 100~MeV, which is convincingly explained by the above "standard" synchrotron radiation reaction limit, $\epsilon_{\rm ph, max} \simeq (9\hbar c/4e^2)\, m_e c^2 \simeq 160$~MeV,  \citep{Guilbert_etal-1983, deJager_etal-1996, Lyutikov-2010, Uzdensky_etal-2011, Komissarov_Lyutikov-2011}.  This indicates that the theoretical reasoning behind this limit is solid and the limit is indeed applicable in real situations, at least under normal circumstances. 

It turns out, however, that this is not the whole story, the actual situation is far more interesting. 
The validity of the above standard radiation reaction limit was recently challenged observationally by the discovery, made by the space-based gamma-ray observatories AGILE and FERMI, of short ($\sim$ 1 day), very intense 100 MeV-1 GeV flares in the Crab Nebula \citep{Abdo_etal-2011, Tavani_etal-2011, Balbo_etal-2011, Striani_etal-2011, Buehler_etal-2012, Buehler_Blandford-2014}. 
For basic energetics reasons, the only viable emission mechanism for the flares is still believed to be synchrotron radiation by PeV electrons in milli-Gauss magnetic fields, but the typical energies of the observed flare photons clearly exceed, by a factor of a few, the "standard" $\lesssim 100$~MeV synchrotron radiation reaction limit \citep{Abdo_etal-2011, Tavani_etal-2011}; see \citep{Buehler_Blandford-2014} for recent review). This paradox thus challenges standard theories of high-energy particle acceleration in relativistic astrophysical plasmas and has lead to an intense theoretical effort aimed at resolving it \citep{Uzdensky_etal-2011, Bednarek_Idec-2011, Komissarov_Lyutikov-2011, Yuan_etal-2011, Clausen-Brown_Lyutikov-2012, Cerutti_etal-2012a,  Bykov_etal-2012, Sturrock_Aschwanden-2012, Lyutikov_etal-2012a, Lyubarsky-2012,  Cerutti_etal-2013, Cerutti_etal-2014a, Cerutti_etal-2014b}. 

One promising idea invokes particle acceleration by magnetic reconnection (\citep{Uzdensky_etal-2011, Cerutti_etal-2012a, Cerutti_etal-2013, Cerutti_etal-2014a, Cerutti_etal-2014b}; see also \citep{Bednarek_Idec-2011}).  The main idea is based on a specific peculiar property of the reconnection process that allows one to circumvent the usual expectation (on which the standard radiation reaction limit is based) that the accelerating electric field $E$ be weaker than the perpendicular magnetic field $B_\perp$ that causes the particle to radiate and hence lose its energy.  Indeed, this expectation is usually well justified almost everywhere in astrophysical plasmas and is related to the applicability of ideal MHD, but intense reconnection layers are precisely the places where ideal MHD reconnection does not apply and hence where one can expect the condition $E < B_{\perp}$ to break down! In fact, the reconnecting magnetic field vanishes exactly at the X-point at the center of a current layer, whereas the electric field there remains finite.  
Thus, one may expect that the Crab flare paradox can be resolved if the required particle acceleration to PeV energies takes place deep inside a reconnection layer, where the magnetic field is weak and so the associated synchrotron radiation reaction force is greatly reduced \citep{Uzdensky_etal-2011, Cerutti_etal-2012a}.  What makes this scenario particularly attractive is that ultra-relativistic particles moving along relativistic Speiser trajectories in a current layer have a natural tendency to focus deeper and deeper into the layer as they gain energy \citep{Kirk-2004, Contopoulos-2007, Uzdensky_etal-2011, Cerutti_etal-2012a}. 
This leads to the formation of discrete highly focused and very intense beams of energetic particles that can be accelerated by the reconnection electric field to energies well above the radiation reaction limit  $\gamma_{\rm rad}^{\rm synch}$ associated with the upstream reconnecting magnetic field $B_0$.  Eventually, these particles escape the low-$B_\perp$ accelerating region and enter a finite-$B_\perp$ region where they quickly radiate their energy in an intense short burst of synchrotron radiation above 100~MeV. The plausibility of this picture, first suggested analytically by~\citep{Uzdensky_etal-2011}, has then been tested in both test-particle (Cerutti et al. 2012a) and fully self-consistent numerical simulations using the radiative relativistic PIC code Zeltron \citep{Cerutti_etal-2013, Cerutti_etal-2014a, Cerutti_etal-2014b}.  This latter study was one of the first (second only to Ref.\citep{Jaroschek_Hoshino-2009} numerical PIC studies of magnetic reconnection that incorporated the radiation reaction force, and also the first to compute the observable radiative signatures (photon spectra and light curves) of reconnection.  

This example illustrates that, whereas there exist important, high-profile astrophysical phenomena where radiation-reaction effects on particle acceleration are expected to play an important role, how exactly these effects play out, in particular, in the case of reconnection-powered synchrotron radiation, is highly non-trivial and extremely interesting.


\subsection{Reconnection-Powered Particle Acceleration in Accreting Black Hole Coronae}
\label{subsec-BH_ADC}

Another important area in high-energy astrophysics where radiation reaction may play an important role in limiting relativistic electron (and perhaps positron) acceleration by reconnection, with potentially important observational consequences, is represented by accretion disks and their coronae in black hole systems, such as galactic X-ray binaries (XRBs) and~AGN.  Here, unlike in pulsar systems, the main radiative mechanism is inverse-Compton (IC) scattering of soft (10-100 eV in AGN and $\sim 1\, {\rm keV}$ in XRBs) accretion-disk photons by the energetic electrons accelerated in coronal reconnection events. This is especially so in bright systems accreting at a significant fraction of the Eddington limit, such as quasars and microquasars (in the high-soft state).  
ADCe in such systems often have Thomson optical depth of order unity and the reconnection layers responsible for the coronal heating and the hard X-ray production is often marginally-collisionless \citep{Goodman_Uzdensky-2008}.  Importantly though, the ambient soft photon field produced by the underlying accretion disk is so intense that the resulting IC radiation reaction is very strong and needs to be taken into account. In particular, it results in an effective Compton-drag resistivity (see \S~\ref{subsec-rad_resistivity}) which,  under some conditions, becomes  greater than the Spitzer resistivity due to electron-ion Coulomb collisions \citep{Goodman_Uzdensky-2008}. 
And, relevant to our present discussion, radiation reaction due to both IC and synchrotron mechanisms can affect the high-energy end of the electron distribution function and hence the observable hard X-ray and gamma-ray emission (e.g., \citep{Khiali_etal-2015a, Khiali_etal-2015b}). 

As discussed above, the relative importance of radiation reaction in reconnection-driven particle acceleration can be assessed by examining the radiative cooling length for electrons at the radiation reaction limit, 
\beq
\ell_{\rm cool} (\gamma_{\rm rad}) = c t_{\rm cool} (\gamma_{\rm rad})  = \gamma_{\rm rad}\, m_e c^3/ P_{\rm rad} (\gamma_{\rm rad}) \sim \beta_E^{-1}\, \rho(\gamma_{\rm rad}) \, , 
\eeq 
and comparing it with other important length scales in the system. 
For typical conditions in ADCe of XRB black holes accreting near the Eddington limit, $\gamma_{\rm rad}^{\rm IC}\, m_e c^2$ can to be of the order of 1000~MeV; thus, there should be virtually no electrons with energies much higher than the proton rest mass (perhaps multiplied by a factor of a few) in these systems.
The corresponding cooling length of these energetic electrons is then comparable to (or perhaps somehwhat larger than) the fiducial proton Larmor radius~$\rho_{i0} \sim m_p c^2/eB_0 = 0.3\, {\rm cm}\, B_7^{-1}$. 
Obviously, this length is much smaller than the typical expected reconnection layer length in~ADC, $L \sim R_g$, where $R_g \equiv GM/c^2 \simeq 1.5 \, {\rm km}\,  M/M_\odot$ is the gravitational radius of a black hole of mass~$M$ and $M_\odot \simeq 2\times 10^{33}\, {\rm g}$ is the solar mass.  For example, for a typical XRB stellar-mass black hole with $M \sim 10 M_\odot$, we have $R_g = 15\, {\rm km}$, and for a typical large super-massive black hole with $M\sim 10^8 M_\odot$, we have $R_g \sim 1.5 \times 10^{8} \, {\rm km} \approx 1\, {\rm AU}$.  

Interestingly, $\gamma_{\rm rad}^{IC}\, m_e c^2 \sim 1000\, {\rm MeV}$ is also comparable to the average dissipated energy per particle, $\bar{\gamma}\, m_e c^2 \sim B_0^2/(16 \pi n_e)$, provided that electrons and ions get comparable amounts of energy and that $\sigma_i \equiv B_0^2/(4\pi n_i m_p c^2) \sim 1$ in the corona.  This in turn means that the above radiation reaction energy limit is more or less comparable to Werner {\it et al}'s "natural" cutoff~$\gamma_{c1} \sim 10\, \bar{\gamma}$ (see \S~\ref{subsec-passive-nonthermal}).  
Thus, provided that that cutoff, discovered in 2D PIC simulations of relativistic pair-plasma reconnection \citep{Werner_etal-2014}, also applies to electron-ion plasmas, we see that $\gamma_{\rm rad}^{IC}$ may in fact be the smallest, and hence the governing, cutoff that limits nonthermal electron acceleration in coronae of real black holes accreting matter at high accretion rates (e.g., of XRBs in the high-soft state). 

The most important potential observational implication of these findings is the prediction that reconnection events in accretion disk coronae of XRB black holes in the high-soft state should be able to produce power-law high-energy IC radiation spectra extending to photon energies on the order of 
$\epsilon_{\rm ph,\, max} \sim \epsilon_{\rm ph,\, soft} (\gamma_{\rm rad}^{IC})^2 \sim 
10^4-10^6 \, \epsilon_{\rm ph,\, soft} \sim 10-1000 \, {\rm MeV}$, 
where we took $\gamma_{\rm rad}^{IC} \sim 100-1000$ and 
$\epsilon_{\rm ph,\, soft} \sim 1\, {\rm keV}$, a typical energy for the dominant radiation emitted by Shakura-Sunyaev \citep{Shakura_Sunyaev-1973} accretion disks around stellar-mass black holes. 

However, because of the high compactness of these systems, most of these high-energy gamma-ray photons probably get absorbed by other photons and create electron-positron pairs before they can escape. This could be the primary process that governs (or at least strongly contributes to) the rate of pair production in black-hole coronae and thus may affect the composition (pairs vs. electron-ion plasma) of black-hole-powered winds and jets. 
Thus, our ability to calculate from first principles the number of electrons accelerated by reconnection to tens
and hundreds of MeV, and hence the number of IC photons at these extreme energies, should give
us an important handle on the efficiency of pair production in these systems --- a fundamental issue
in black-hole astrophysics.

\section{Reconnection with Radiative cooling}
\label{sec-rad_cooling}

One of the most important effects that radiative drag force can have on reconnection, and one that often comes into play first in various astrophysical contexts, is its effect on the random thermal motions of average, run-of-the-mill particles in the reconnection layer. We call this effect {\it radiative cooling}.  Here we are particularly interested in the case of prompt radiative cooling, in which the characteristic cooling time $t_{\rm cool}$ of the hot plasma energized by the reconnection processes is shorter than or comparable to the characteristic time that a given fluid element spends inside the reconnection layer, which is typically the Alfv\'en transit time along the layer, $\tau_A= L/V_A$. 
In this regime, which we will call the strong radiative cooling regime, radiative energy losses become important in the overall energy balance of the reconnection process and need to be taken into account. The opposite case was considered in~\S~\ref{sec-passive-rad-signs}.

It is easy to estimate that radiative cooling is not important in most traditional (solar-system) applications of reconnection, i.e., in the environments of solar flares, the Earth magnetosphere, and tokamak fusion devices and dedicated laboratory experiments designed to study reconnection under controlled laboratory conditions.  Because of this, there has been relatively little work done on incorporating radiative cooling effects into reconnection models.  However, when one tries to think about reconnection in various astrophysical contexts, one often finds, by doing simple estimates, that if magnetic reconnection happens in these environments, it has to take place in the strong radiative cooling regime. 
This realization has lead to an increased interest in radiative magnetic reconnection in the high-energy astrophysics community, especially in recent years 
~\citep{Dorman_Kulsrud-1995, Lyubarsky-1996, Jaroschek_Hoshino-2009, Giannios_etal-2009, Nalewajko_etal-2011, Uzdensky-2011, Uzdensky_McKinney-2011, McKinney_Uzdensky-2012, Takahashi_Ohsuga-2013, Takahashi_Ohsuga-2015, Uzdensky_Spitkovsky-2014}.
The importance of radiative cooling effects on reconnection have also been recognized in the context of reconnection in the solar chromosphere \citep{Steinolfson_vanHoven-1984, Leake_etal-2013, Ni_etal-2015}.

\subsection{Reconnection with Radiative Cooling: Optically Thin Case}
\label{subsec-rad_cooling-thin}

The first key step to study the effects of strong cooling on reconnection in a systematic way was made recently by Uzdensky \& McKinney (2011) \citep{Uzdensky_McKinney-2011}, who developed a simple but self-consistent Sweet--Parker-like model for a non-relativistic resistive-MHD reconnection layer subject to strong optically thin cooling.  
It is of course understood that, just like the original Sweet--Parker model, this model should not be expected to provide a complete description of reconnection in real large astrophysical systems, which are often subject to a host of other effects, such as ambient turbulence \citep{Lazarian_Vishniac-1999}, secondary current-layer instabilities such as the plasmoid instability \citep{Loureiro_etal-2007},  and collisionless effects (e.g., \citep{Birn_etal-2001}), just to name a few.
For these reasons the Uzdensky \& McKinney (2011) model of reconnection with radiative cooling should only be 
viewed as a toy model which, however, brings out several important physical insights into the problem and provides a useful fundamental building block for future studies. 

The main idea of this model was that since radiative cooling limits the rise of the plasma temperature in the layer, in the absence of a guide magnetic field, the plasma inside the layer has to compress in order to maintain the pressure balance with the upstream magnetic pressure. By analyzing carefully the balance between ohmic heating and radiative and advective cooling, together with the cross-layer pressure balance and with the equation of motion along the layer, Ref.~\citep{Uzdensky_McKinney-2011} obtained estimates for the key parameters characterizing the reconnecting system: the plasma compression ratio, the reconnection rate, and the layer thickness, in terms of the general parameters of the radiative cooling function. 
It was found that the reconnection rate, in the case with no guide field, is enhanced relative to the non-radiative case and that the layer thickness is reduced due to the cooling-related compression; a strong guide field, however, suppresses this effect by preventing strong compression.  In addition, for the specific case of Spitzer resistivity~$\eta_{\rm Sp}$, reconnection is sped up (with or without a guide field) by radiative cooling even further due to the strong inverse scaling of the resistivity with temperature, $\eta_{\rm Sp} \sim T^{-3/2}$. 
Furthermore, several specific astrophysically-important radiative mechanisms (bremsstrahlung, cyclotron, and inverse Compton) were considered and the conditions for strong-cooling regime were formulated for each one of them.  The theory lead to specific expressions for the reconnection rate and to the prediction of a cooling catastrophe behavior for the case of strong bremsstrahlung cooling. 
Although this study~\citep{Uzdensky_McKinney-2011} focused mostly on optically thin case, many of its ideas, concepts, and conclusions should be valid more broadly; however, analyzing reconnection dynamics in the optically-thick case requires approaching the problem as a radiative-transfer problem, as we discuss in~\S~\ref{subsec-rad_cooling-thick}.


An interesting astrophysical example of reconnection where strong radiative cooling is important is reconnection in the magnetospheres of gamma-ray pulsars (e.g., the Crab), at distances comparable to, but also perhaps somewhat larger than the light cylinder (LC) radius (\citep{Uzdensky_Spitkovsky-2014}; see also \citep{Lyubarsky-1996, Arka_Dubus-2013}). 
The rotating pulsar magnetosphere naturally develops an equatorial current sheet beyond the light cylinder, somewhat similar to the heliospheric current sheet, and magnetic reconnection in this current sheet can dissipate a nontrivial fraction of the overall pulsar spin-down power within a few LC radii.  

In some rapidly rotating pulsars, the reversing magnetic field just beyond the light cylinder is so strong (e.g., of order 1 MG for the Crab pulsar) that prompt synchrotron cooling of the heated plasma in the layer inevitably becomes important; it controls the energetics of reconnection and may result in production of observed strong pulsed GeV $\gamma$-ray emission \citep{Lyubarsky-1996, Uzdensky_Spitkovsky-2014, Arka_Dubus-2013}. 

In particular, by combining the conditions of the pressure balance across the current layer (reconnection in pulsar magnetosphere is expected to take place without a guide field) and of the balance between the heating by magnetic energy dissipation and synchrotron cooling, one can obtain simple estimates for key physical  parameters of the layer's plasma, such as the temperature, density, and layer thickness, in terms of the reconnecting upstream magnetic field~$B_0$ \citep{Uzdensky_Spitkovsky-2014}. Specifically, one expects the plasma to be heated to roughly the radiation reaction limit $\theta_e  \equiv T_e/m_e c^2 \sim \gamma_{\rm rad}^{\rm synch}$ (see Eq.~\ref{eq-gamma_rad-synch}) and thus to be compressed to the density $n_e \sim B_0^2/(16\pi \gamma_{\rm rad}^{\rm synch} m_e c^2)$ (in the comoving frame of the relativistic pulsar wind). 
The corresponding thickness of the small elementary inter-plasmoid current layers is then expected to be comparable to the relativistic Larmor radius of these particles, $\delta \sim \rho_L(\gamma_{\rm rad}^{\rm synch}) = \gamma_{\rm rad}^{\rm synch} \rho_0$.  For the particularly important case of the Crab pulsar, one finds: 
$T_e \sim 10^4 \, m_e c^2 \sim 10\, {\rm GeV}$, $n_e \sim 10^{13}\, {\rm cm^{-3}}$, and $\delta \sim 10\, {\rm cm}$.  After accounting for the bulk Doppler boosting due to the pulsar wind (with a Lorentz factor of order 100), the synchrotron and inverse-Compton emission from the reconnecting current sheet may plausibly explain the observed Crab's pulsed high-energy (GeV) and VHE ($\sim 100$ GeV) radiation, respectively, while the rapid motions of the secondary plasmoids in the large-scale current layer may contribute to the production of the pulsar radio emission \citep{Uzdensky_Spitkovsky-2014}. 


In addition to astrophysical applications, magnetic reconnection in the strong optically thin radiative cooling regime may soon be within reach to laboratory studies utilizing powerful modern laser plasma facilities, such as Omega EP and NIF (Uzdensky et al. 2016, in prep.).  The main cooling mechanism in these experiments is collisional bremsstrahlung, perhaps augmented by atomic-line cooling, depending on the plasma composition.  Since the bremsstrahlung cooling rate scales strongly with the plasma density (as $n_e^2$) but only weakly with the temperature (as $T^{1/2}$), in order to reach the desired radiative regime, it is advantageous to configure the laser-target setup towards a higher density, a lower temperature, and a larger illuminated area.  In addition, the role of radiative cooling is enhanced if one uses targets made of high-$Z$ materials, such as copper and gold.  Overall, preliminary estimates indicate that the strong radiative cooling regime is reachable on NIF and perhaps even on Omega EP when using gold targets. 
Interestingly, some of the physical parameters achievable in these laser-plasma experiments, e.g., magnetic field strengths, densities, characteristic kinetic plasma length scales, are not that different from the values expected in, e.g., BH accretion disk coronae in XRBs. This points to tantalizing potential prospects of studying in the lab the magnetic reconnection processes in the regimes relevant to these astrophysical environments.

\subsection{Reconnection with Radiative Cooling: Optically-Thick Case}
\label{subsec-rad_cooling-thick}

In the optically-thick radiative cooling case, i.e., when the optical depth $\tau$ {\it across} the layer is large, 
a self-consistent treatment of radiation calls for serious modifications to our overall theoretical approach to the reconnection problem. 
Specifically, one has to view the reconnection problem in this case essentially as a radiative transfer problem \citep{Uzdensky-2011}.  
The current layer develops a photosphere, with a photospheric surface temperature, $T_{\rm ph}$, which in a steady state is related to the temperature $T_0$ at the center of the reconnection layer via 
$T_0^4/T_{\rm ph}^4 = \tau \gg 1$. 
Furthermore, in the strong-cooling regime the basic steady-state energy balance between the Poynting flux entering the layer from upstream with the reconnection inflow, 
$S  = (c/4\pi)\,  E_{\rm rec} B_0 = v_{\rm rec}\, B_0^2/4\pi = c\, \beta_{\rm rec} B_0^2/4\pi$, 
and the outgoing radiative flux emitted by the photosphere, 
$F_{\rm rad} = \sigma_{\rm SB} T_{\rm ph}^4$, 
where 
$\sigma_{\rm SB} = \pi^2 k_B^4/60 \hbar^3 c^2 \simeq 5.67 \times 10^{-5}\, {\rm erg \,cm^{-2}\, s^{-1} \, K^{-4}}$ 
is the Stefan-Boltzmann constant, 
determines the photospheric temperature in terms of the reconnecting magnetic field:
\beq
T_{\rm ph} = \biggl[ {c\,\beta_{\rm rec}\over{\sigma_{\rm SB}}} \, {{B_0^2}\over{4\pi}} \biggr]^{1/4} \, ,
\label{eq-opt_thick-T_ph}
\eeq
and hence the central layer temperature for a given~$\tau$: 
\beq
T_0 = \tau^{1/4} T_{\rm ph} = 
\biggl[ \tau {c\,\beta_{\rm rec}\over{\sigma_{\rm SB}}} \, {{B_0^2}\over{4\pi}}  \biggr]^{1/4} \, .
\label{eq-opt_thick-T_0}
\eeq

Next, if there is no guide field and if the pressure inside the layer is dominated by the gas pressure, $P_0 = 2 n_{e,0} k_B T_0$, then one can use the condition of pressure balance across the layer between $P_0$ inside the layer and the combined magnetic plus plasma pressure in the upstream region outside the layer, $(1 + \beta_{\rm up})\, B_0^2/(8\pi)$, where $\beta_{\rm up}$ is the upstream plasma-$\beta$ parameter, to obtain the central plasma density: 
\beq
n_{e,0} = {B_0^2\over{16\pi}}\, {{1+\beta_{\rm up}}\over{k_B T_0}} \, .
\label{eq-opt_thick-n_0}
\eeq

The expressions (\ref{eq-opt_thick-T_ph})-(\ref{eq-opt_thick-n_0}) govern the basic thermodynamics of a reconnection layer subject to strong radiative cooling in the optically thick regime. 


\subsection{Optically-Thick Current Layer: Radiation Pressure Effects}
\label{subsec-rad_pressure}

However, in some important astrophysical phenomena, the reconnecting magnetic field is so strong and hence the total plasma energy density in the layer is so high that the pressure is dominated by radiation pressure. 
For an optically thick layer, we can assume thermal black-body radiation pressure: 
$P_{\rm rad,0} = a T_0^4/3$, where 
$a = 4\sigma_{\rm SB}/c \simeq 7.57 \times 10^{-15}\, {\rm erg\, cm^{-3}\, K^{-4}}$ 
is the radiation constant. 
In this case, the cross-layer pressure balance does not involve the plasma density and instead yields, in combination with the steady-state energy balance 
$S=c\,\beta_{\rm rec} B_0^2/4\pi = 
F_{\rm rad} = \sigma_{\rm SB} T_{\rm ph}^4 = \tau^{-1}\,\sigma_{\rm SB} T_0^4$, 
a simple but important relationship between the optical depth and the reconnection rate \citep{Uzdensky-2011}: 
\beq
\tau \, \beta_{\rm rec} = {3\over 8} \, .
\eeq

The validity of this expression is limited by a few assumptions, namely, by the steady-state assumption and by the assumption of strong cooling. The latter, in particular, implies that the radiative diffusion time across the layer, 
$t_{\rm diff}  \sim \tau \delta/c$ is much shorter than the (Alfv\'enic) advection time $\tau_A  \sim L/V_A$ along the layer, and this imposes a certain condition on the layer's aspect ratio relative to the optical depth: 
$ L/\delta > \tau V_A/c$.

\section{Radiation Drag on Fluid Flow}
\label{sec-rad_drag}

Another fluid-level manifestation of the radiative drag force is its direct braking effect on bulk plasma motions expected in a reconnecting system. Thus, in contrast to radiative cooling, which affects the {\it thermodynamics} of the reconnection process, the radiation effects that we consider in this section influence the {\it dynamics} and {\it electrodynamics} of reconnection.  Here it is convenient to distinguish two aspects of such action: \\
{\bf (1)} radiative friction on the current-carrying charged particles moving in the out-of-plane direction and responsible for carrying the current in the reconnection layer, resulting in an effective {\it radiative resistivity}; \\
{\bf (2)} radiative friction slowing down the {\it reconnection outflows} in the direction along the reconnecting magnetic field (the outflow direction). 

Although the actual underlying physical mechanism behind these two effects is the same, for practical reasons it is convenient to consider them separately  because they play different roles in changing the reconnection dynamics. 

\subsection{Radiative Resistivity}
\label{subsec-rad_resistivity}

When electrons carrying the electric current in a reconnecting current layer drift through an external radiation field, or perhaps radiate themselves via, e.g., synchrotron radiation, the radiation drag force on the electron flow may produce an effective radiative resistivity, $\eta_{\rm rad}$.  

In particular, for non-relativistic electrons, e.g., in applications such as accretion disks and ADCe, equating the radiation drag force given by Eq.~(\ref{eq-f_rad-IC}) with the accelerating electric force, 
one finds the steady-state drift velocity $<{\bf v}_e> = -\, (3/4) (ce/\sigma_T U_{\rm rad}) \, {\bf E}$.  
These electrons thus  carry an electric current of ${\bf j}_e = - e n_e <{\bf v}_e> =  (3/4)\, (cn_e e^2/\sigma_T U_{\rm rad}) \, {\bf E}$, which corresponds to an effective electron contribution to the electric conductivity of  
\beq
\sigma_{e,\rm rad}^{\rm IC}  = {3\over 4} \, {{cn_e e^2}\over{\sigma_T U_{\rm rad}}} \, .
\label{eq-sigma_e_IC}
\eeq

It is interesting to note that, as was discussed by \citep{Goodman_Uzdensky-2008}, strictly speaking, Compton drag does not change the steady-state resistivity in an electron-ion plasma; perhaps counter-intuitively, the resistivity actually just remains the Spitzer collisional resistivity. This is because radiation drag essentially affects only the electrons but not the ions (because the Thomson cross-section scales as the inverse square of the particle mass). Therefore, even if the electrons are greatly slowed down by the radiation field, the ions can eventually (if long enough time scales and length scales are available) get accelerated by the applied electric field to carry the necessary current.  In many important astrophysical applications, however, including reconnection in BH ADCe, one is interested in processes that take place on such short length scales that the ions may not enough range to get accelerated to the Coulomb-collision-limited steady-state drift velocity.  In such situations, one can ignore the ion current and hence cast the effective Ohm's law in terms of an effective Compton-drag resistivity, or a Compton magnetic diffusivity given by 
\beq
\eta_{\rm IC} = {c^2 \over{4\pi \sigma_{e,\rm rad}^{\rm IC}}} = 
{1\over{3\pi}} \, {{c \sigma_T U_{\rm rad}}\over{n_e e^2}} \, .
\label{eq-eta_IC}
\eeq

In a pair plasma, of course, both electrons and positrons are subject to radiative drag equally; hence, their conductivities are both given by~(\ref{eq-sigma_e_IC}) (in the non-relativistic regime) and therefore the total radiative resistivity equals one half of the value given by Eq.~(\ref{eq-eta_IC}).

As we discussed in \S~\ref{subsec-BH_ADC}, in astrophysical applications such as coronae of accretion disks of black holes accreting at a large fraction of the Eddington rate~$L_E$, the ambient radiation field, with a radiation energy density $U_{\rm rad} \sim L_E/4\pi R^2 c $, where $R \simeq 10 R_g$ is the characteristic size of the bright inner part of the accretion disk, is very intense.  
Under such conditions, the resulting effective radiative resistivity can be quite high and may dominate over the Spitzer resistivity due to classical Coulomb collisions. 
It can then seriously affect the reconnection processes that are believed to be responsible for coronal heating and for powering the observed hard X-ray emission from these sources.  In particular, enhanced radiative resistivity may alter the analysis of whether the global reconnection layer is in the collisional or collisionless regime \citep{Goodman_Uzdensky-2008} and may thus affect (reduce) the reconnection rate and the hierarchy of secondary plasmoids emerging in the reconnection layer.  Needless to say, however, the regime where IC effective resistivity is important probably also implies that radiative cooling is important as well, which may actually speed up reconnection (see section~\ref{subsec-rad_cooling-thin}). In this case, it is not yet clear what the overall combined effect of radiation (cooling plus resistivity) on the reconnection rate is. 

In the relativistic case, the Compton-drag resistivity was calculated by \citep{van_Oss_etal-1993}. 
An important point to keep in mind when considering effective resistivity in the relativistic case is that electric current depends only on the 3-velocity of charge carriers and not on their Lorentz factor. Thus, as long as the particles are ultra-relativistic (and thus travel nearly at the speed of light), the current density they can carry is limited by $e n_e c$, independent of the electric field.  The effect of radiation drag on the resistivity in this case is diminished. 

In addition to black-hole accretion disks and their coronae, radiative resistivity due to various radiative mechanisms may also potentially play a role in reconnection processes in a number of other high-energy astrophysical systems, e.g., in the magnetospheres of pulsars \citep{Uzdensky_Spitkovsky-2014}, magnetars  \citep{Uzdensky-2011}, and GRBs~\citep{McKinney_Uzdensky-2012}.

\subsection{Radiative drag on reconnection outflow}
\label{subsec-rad_drag-outflow}

Finally, let us discuss the effects of radiative drag on the bulk outflow from the reconnection region. This outflow is an important aspect of the reconnection process; its main role is to evacuate from the reconnection layer the plasma that flows into the layer bringing in fresh unreconnected magnetic flux, and thus to make room for more plasma to enter.  The outflow thus represents an important element of the overall stagnation-flow pattern around the magnetic X-points. The outflow is driven by a combination of the pressure gradient force and the magnetic tension force associated with reconnected magnetic field lines. It usually represents the fastest motion found in a reconnecting system, with a speed on the order of the Alfv\'en speed and hence significantly higher than the reconnection inflow velocity.  Importantly, the outflow can usually be described roughly as an ideal-MHD motion as it involves the electrons and ions moving together in the same direction%
\footnote{Strictly speaking in weakly collisional plasmas this is not quite correct since the electron and ion outflow patterns are somewhat different, which  results in an in-plane current circulation responsible for the quadrupole out-of-plane magnetic field.}. 

In a number of astrophysical applications, including TeV flares in blazar jets \citep{Nalewajko_etal-2011} and black-hole accretion disks, GRB jets, and others, the Compton drag due to ambient radiation may have a substantial effect on the reconnection outflow. It can slow down the outflow, choking the motion of plasma through the reconnection system and thereby reducing the reconnection rate in a manner similar to the effect of a large viscosity. 
For example, \citep{Takahashi_Ohsuga-2013, Takahashi_Ohsuga-2015}, using relativistic resistive MHD simulations that included optically-thick radiation effects, reported that radiative drag on the reconnection outflow lead to a reduction of the reconnection rate for Petschek-like relativistic reconnection. 

In addition, as the ambient isotropic radiation field exerts a braking Compton-drag force on the plasma flow, it also extracts its energy and can, under certain circumstances, convert a noticeable fraction of the bulk kinetic energy of the outflow into radiation beamed in the outflow direction.  


To illustrate the effect of radiative braking of the outflow on the reconnection rate, let us consider a simple, Sweet-Parker-like toy model of a laminar non-relativistic incompressible resistive-MHD reconnection problem. 
For illustration, we will only take into account the Compton-drag force in the outflow fluid equation of motion and will ignore other radiative effects such as radiative cooling and resistivity.  Since the inflow is generally much slower than the outflow, the effect of the radiative drag on the inflow can also be neglected. We will also ignore the guide magnetic field. 
Furthermore, we will focus on an extreme case where radiative drag dominates over the plasma inertia in establishing the ultimate outflow velocity.  The model is then somewhat similar to the analysis of resistive Hall-MHD reconnection in Ref.~\citep{Uzdensky-2009}. 

For definiteness, let us choose a system of coordinates with $x$ being the direction of the reconnecting magnetic field (and hence of the reconnection outflow), $y$ being the direction across the layer, and $z$ being the ignorable direction. 
Then, ignoring the plasma inertia in the outflow ($x$) equation of motion, the outflow velocity $u_x$ is governed by the balance between the outward pressure gradient force $-dP/dx$ (magnetic tension may give a comparable contribution but we ignore it here for simplicity) and the Compton-drag force (per unit volume),  
$- (4/3) \, n_e \sigma_T U_{\rm rad} \, u_x/c$. This yields the following estimate for the final outflow speed at the end of the layer of length~$L$: 
\beq
u_{\rm out} \sim c \, {\Delta P \over L} \, {1\over{n_e \sigma_T U_{\rm rad}}} = 
c \, {\Delta P \over{\tau_T\, U_{\rm rad}}} \, , 
\eeq
where $\tau_T \equiv n_e \sigma_T L$ is the Thomson optical depth along the layer. 
The drop $\Delta P$ of the plasma pressure along the layer can be estimated, as is done in the traditional Sweet-Parker model, by using the condition of pressure balance across the layer, $P_0 = P^{\rm up}  + B_0^2 /8\pi$, 
and ignoring the variation of the upstream plasma pressure $P^{\rm up}$ along the layer. 
Thus, $\Delta P \simeq B_0^2 /8\pi$ and we get 
\beq
u_{\rm out} \sim c \, {{B_0^2}\over{8\pi \tau_T\, U_{\rm rad}}}  = 
c \, \tau_T^{-1}\, {{U_{\rm magn}}\over{U_{\rm rad}}}\, .
\eeq

One can see that, since we assumed the outflow to be non-relativistic, $u_{\rm out} \ll c$, 
this result requires that the radiation energy density times the optical depth be sufficiently large compared to the magnetic energy density, i.e.,  $\tau U_{\rm rad} \gg U_{\rm magn}$, a condition that is indeed satisfied, for example, in the inner parts of black-hole accretion disks. 
Furthermore, since in this model we neglected the plasma inertia compared to the radiation drag, we must also require that $u_{\rm out} \ll V_A = B_0\, (4\pi \rho)^{-1/2}$.  This, in turn, imposes an even more stringent condition than $u_{\rm out} \ll c$, namely, $\tau U_{\rm rad} \gg (U_{\rm magn}\, \rho c^2)^{1/2}$, which, however, can also be satisfied inside black-hole accretion disks. 

The rest of the reconnection problem analysis is the same as in the classical Sweet-Parker model. 
Employing the incompressibility condition, $\delta u_{\rm out} = v_{\rm rec} L$, 
and the steady-state resistive magnetic induction equation: 
$ \delta v_{\rm rec} = \eta$, where $v_{\rm rec}$ is the reconnection inflow velocity, $\delta$ is the layer thickness, and $\eta$ is the magnetic diffusivity (which may, in general, be due to both Coulomb collisions and radiative resistivity), one obtains the usual Sweet-Parker scaling for the reconnection rate and for the aspect ratio: 
\beq
{\delta\over L} \sim {v_{\rm rec} \over{u_{\rm out}}} \sim S_{\rm rad}^{-1/2} \, , 
\eeq
where, however, the radiation-controlled outflow velocity $u_{\rm out}$ replaces the Alfv\'en speed in the effective Lundquist number, i.e., 
\beq
S_{\rm rad} \equiv  L \, {u_{\rm out}\over \eta} \, . 
\eeq

This model, although highly simplified, may provide a useful building block in constructing a more complete theoretical picture of magnetic reconnection in certain radiation-rich astrophysical environments, for example, in the context of high accretion rate black-hole accretion flows in XRBs and~AGNs. 

\section{Other radiation effects in optically thick plasmas: 
radiation pressure, radiative viscosity and hyper-resistivity, and pair creation}
\label{sec-other}


In systems with non-negligible optical depth across the layer some of the photons produced by the reconnection process do not promptly leave the system but may interact with the particles in the layer again by scattering or absorption (or pair creation at higher energies, see below).  This interaction can lead to additional effects, such as radiation pressure and radiative viscosity, both of which can affect the reconnection dynamics. 

In particular, if the layer is optically thick to scattering, then radiation pressure $P_{\rm rad}$ enters the pressure balance across the layer:
\beq
P_{\rm gas} + P_{\rm rad} + B^2 / 8\pi = {\rm const} \, .
\eeq
This implies that the plasma pressure at the center of the layer does not need to increase as much as in the case without radiation pressure in order to balance the outside magnetic pressure.  For example, if the optical depth is large enough for radiation to reach local thermal equilibrium with the plasma at the local plasma temperature~$T$, then $P_{\rm rad} = a T^4/3$ and hence the pressure balance becomes 
\beq
2 n k_B T + aT^4/3 + B^2 / 8\pi = {\rm const} \, .
\eeq
Therefore, the temperature in the layer can be lower than in the case without radiation pressure. 
The thermodynamic structure of the layer in this case is determined by the radiative transfer problem across the layer.
If, however, the optical depth is modest, $\tau \lesssim 1$, then the effects of radiation pressure are reduced by a factor of $\tau$, but can still be significant under some circumstances. 

Because of the very steep dependence of the radiation pressure on temperature, we can see that radiation pressure effects are important mostly only in very hot environments. In addition, since the optical depth must be high enough to ensure a good coupling of the radiation pressure to the plasma, the density must also be high. From this one can deduce that radiation pressure effects on reconnection are expected to be important mostly in high-energy-density systems.  Most notable astrophysical examples of such systems include the inner parts of black-hole and neutron-star accretion disks in binary systems; magnetospheres of normal neutron stars, e.g., X-ray pulsars (with "normal" magnetic fields of about $10^{12}\, {\rm G}$), and magnetars in, e.g., SGR systems (with magnetic fields of order $10^{15}\, {\rm G}$); and central engines of supernovae (SNe) and~GRBs. 


In addition to the radiation pressure effects in a reconnection layer of non-negligible optical depth, the momentum extracted by radiation from reconnection outflow (see the discussion of radiative braking in \S~\ref{subsec-rad_drag-outflow}) can be deposited back to the plasma in other parts of the layer, which results in an effective viscosity mediated by the photons.  This radiative viscosity is then expected to  affect the basic reconnection dynamics in a manner similar to the usual collisional viscosity caused by the thermal motions of electrons and ions: it should lead to broadening of the layer and to decreasing the reconnection rate. 

Likewise, the momentum extracted by radiation from current-carrying electrons  in the current layer (which leads to a Compton-drag radiative resistivity, see \S~\ref{subsec-rad_resistivity}) can be deposited to other current-carrying electrons elsewhere in the layer if the optical depth is not negligible. This essentially spreads the electric current, making the layer broader, an effect that can be described as a result of a radiative hyper-resistivity proportional to the optical depth (for $\tau< 1$).  
Interestingly, this hyper-resistivity should only work in electron-ion plasmas; in pair plasmas, since photons can be scattered or absorbed by both electrons and positrons (which drift in opposite directions), radiative hyper-resistivity just gives way to an enhanced radiative resistivity. 


In the most extreme astrophysical systems, such as the magnetospheres of magnetars in SGR systems and GRB and SN central engines, the reconnecting magnetic field $B_0$ exceeds the quantum critical magnetic field 
$B_Q \equiv \alpha_{\rm fs} B_{\rm cl} = m_e^2 c^3/e \hbar \simeq 4.4 \times 10^{13} \, {\rm G}$. Then, the dissipated magnetic energy density is so high that the pressure of the heated plasma inside the reconnection layer becomes dominated by the radiation pressure and, furthermore, the resulting radiation temperature becomes relativistic, i.e., $T_0 \sim m_e c^2 \, (B_0/B_Q)^{1/2}$ \citep{Uzdensky-2011}.  In this case, prodigious pair production inevitably results and the current layer gets quickly "dressed" in an optically thick and dense pair coat.  Once again, the problem of determining the thermodynamic structure across such a dressed layer becomes a radiative transfer problem, but one in which the pair density at any location in the layer below its pair-creation photosphere is determined by the local thermodynamic equilibrium.  
In particular, for reconnecting magnetic fields that are not just higher, but much higher than~$B_Q$, e.g., in magnetar systems, one expects ultra-relativistically hot plasma, $T \gg m_e c^2$; and under these conditions the pair density scales as~$T^3$ and the pair contribution to the total pressure becomes comparable to the radiation pressure \citep{Uzdensky-2011}.  Magnetic reconnection in this very exotic regime may be the mechanism behind some of the most spectacular, energetic phenomena in the Universe --- giant SGR flares, releasing huge amounts of energy in the form of gamma-rays in just a fraction of a second \citep{Uzdensky-2011, Uzdensky_Rightley-2014}.  This regime can be regarded as the most extreme case of radiative reconnection, because all of the radiative effects discussed in this article --- radiation-reaction limits on particle acceleration, strong radiative cooling, radiation pressure, Compton-drag resistivity, etc. --- are active in this case.

\section{Conclusions and Outlook}
\label{sec-conclusions-outlook}

This Chapter presented a review of the physics of radiative magnetic reconnection and its applications to various astrophysical phenomena.  Traditional reconnection research, motivated by applications to relatively low-energy-density solar-system and laboratory plasma environments, has historically ignored the possible effects and observational signatures of radiation.  In many astrophysical reconnecting systems, however, various radiation effects exert an important influence on the dynamics and energetics of the reconnection process, as well as on the associated nonthermal particle acceleration. These effects ultimately stem from the radiation reaction force on individual particles, which is  directly related to the rate of energy losses suffered by the particle (i.e., the particle's radiative power). 
Since the radiative power is often proportional to the energy density of the external agent field that causes the particle to radiate (e.g., magnetic energy density for synchrotron radiation and ambient radiation energy density for inverse-Compton radiation), we see that the relative importance of the radiation reaction force in the particle equation of motion can usually be traced to the high energy density in astrophysical systems of interest, combined with their large size. 
The main radiation mechanisms involved in high-energy astrophysical reconnection, especially in relativistic systems, are cyclo/synchrotron radiation, curvature radiation, and inverse-Compton scattering.  In addition, bremsstrahlung radiation and pair creation can play a role under some circumstances.  

The radiation reaction force can manifest itself via several different radiative effects, the relative importance of which depends on the particular astrophysical context.  The first radiative effect that comes in at lowest energy densities is the radiation-reaction limit on relativistic particle acceleration. This is a purely kinetic effect, it is due to the fact that for relativistic particles the radiative power, and hence the radiation reaction force, grow rapidly with the particle's Lorentz factor~$\gamma$. This means that the radiation back-reaction first affects the most energetic particles, while leaving lower-energy particles less affected. This necessitates a kinetic treatment.  One of the most prominent astrophysical examples where this effect has to be taken into account in considering magnetic reconnection is the Crab pulsar wind nebula, in particular in relation to the recently discovered short and bright gamma-ray (hundreds of MeV) flares that seem to require extreme particle acceleration to PeV energies, overcoming the synchrotron radiation reaction limit.  Another important astrophysical example is found in reconnection events powering coronal heating and hard-X-ray emission in accretion disk coronae of black holes, e.g., in galactic X-ray binaries and active galactic nuclei. Here, inverse-Compton radiation drag due to the intense ambient soft photon field emitted by the underlying accretion disk imposes interesting upper limits on the electron acceleration. 

Most of the other radiative effects acting in high-energy astrophysical reconnection can be described as fluid-level effects; they affect not just a select few highest energy particles but most of the particle population; thus, they  seriously affect the overall dynamics and energy budget of a reconnection process.  Correspondingly, these effects require very high energy densities, which implies that they usually become important for reconnection events happening close to the central compact object, such as a neutron star or a black hole.  Just to organize our thinking, we can categorize the radiative effects on reconnection according to the different components of the particle motion that are being affected by the radiative drag. 
Thus, radiative drag on random, "thermal', particle motions, especially in the direction across the current layer, effectively leads to radiative cooling, which is reviewed in \S~\ref{sec-rad_cooling}.  It may lead to a substantial plasma compression and speed up the reconnection process. Radiative cooling is important in systems such as reconnecting equatorial current sheet in a pulsar magnetosphere just outside the pulsar light cylinder, perhaps powering the observed pulsed high-energy gamma-ray emission (synchrotron cooling, see \S~\ref{subsec-rad_cooling-thin}); 
inner parts of accretion disks and accretion disk coronae of black hole systems (inverse-Compton cooling, see ~\S~\ref{subsec-BH_ADC}); 
reconnection events in magnetospheres of magnetars, perhaps powering giant gamma-ray flares in Soft Gamma Repeaters; 
and in relativistic jets of gamma-ray bursts. 

Next, radiative drag on the bulk collective motions of electrons (and perhaps positrons) may result in: 
({\it i}) effective radiative (Compton drag) resistivity for the flow of electrons carrying the main electric current in the reconnection current layer, important, e.g., in accreting black hole coronae, magnetar magnetospheres, and central engines of supernovae and gamma-ray bursts; 
and 
({\it ii}) effective braking of the plasma outflow from the reconnection layer, potentially slowing down the reconnection process; this effect has been explored in the context of TeV flares in blazar jets but may also be important in a number of other systems, including accretion disks around black holes and neutron stars. 

Whereas most of the above-mentioned radiative effects can operate in optically thin plasmas, there are some radiation effects that take place in optically thick reconnecting systems. In particular, this may occur at very high plasma densities and energy densities, found, e.g., in systems like the central engines and jets of gamma-ray bursts, magnetar magnetospheres, and perhaps central parts of black hole accretion disks.  Reconnection layers in these environments  may become optically thick, which allows the photons emitted by the energetic particles in the layer to interact with the layer particles again. This secondary interaction opens up avenues for additional radiative effects, namely, radiation pressure and effective radiative viscosity (see \S~\ref{sec-other}). Furthermore, since most of these plasmas are relativistically hot and compact, there are many gamma-ray photons above the pair-production threshold, which makes a copious pair production not only possible but often inevitable. Intense pair production, in turn, further increases the optical depth and plasma collisionality of the reconnection layer. 

Finally, apart from its possible active role in influencing reconnection dynamics and energetics, radiation emitted by a reconnection layer also plays a role of an important (and, quite often in astrophysics, the only) {\it diagnostic tool} that we can use to study remote astrophysical systems.  This applies not only to all of the above-mentioned radiative reconnection systems, but also, arguably, to most astrophysical systems where we believe magnetic reconnection takes place, even when it acts as a purely passive tracer. 
For this reason, it is particularly important to develop theoretical and computational tools that will enable us to predict, calculate potentially observable radiative signatures of a reconnection process. 


One should expect continuing rapid development of the field of radiative magnetic reconnection in the next few years. 
This optimistic outlook for accelerating progress in this exciting new frontier of plasma astrophysics is justified by the convergence of several factors.  First, there is a strong and growing astrophysical motivation for its serious development, based on the increasing recognition by the broad astrophysical community of the importance of magnetic reconnection as a potent mechanism for plasma heating, nonthermal particle acceleration, and high-energy radiation production in numerous astrophysical phenomena.  This leads to an increased interest among astrophysicists in magnetic reconnection in general; however, as argued in this Chapter, in many, if not most, of the astrophysical phenomena of interest reconnection inevitably takes place in the radiative regime, in which  prompt radiative energy losses materially affect the process. In addition, the need to connect reconnection theory to observations, by developing the capability to calculate observable radiative signatures, also contributes to the astrophysical motivation. 

The second fundamental reason for expecting rapid progress in radiative reconnection is the emerging ability to study this reconnection regime in the lab, namely, by using modern high-energy-density laser-plasma facilities, such as Omega EP and NIF.  By using high-$Z_{\rm eff}$ target materials such as gold, it should be possible to achieve a reconnection regime where bremsstrahlung and perhaps atomic-line radiative cooling become important.  In addition, powerful Z-pinch facilities (such as Imperial College's MAGPIE) could also potentially be adapted to laboratory studies of radiative HED reconnection.  All these new experimental capabilities that are now becoming available can potentially provide a valuable research tool, a testbed for validating theories and numerical models of radiative reconnection, and also perhaps lead to completely new, unexpected discoveries. 

Finally, current and future progress in radiative reconnection is greatly facilitated by the appearance of new computational tools, coupled with analytical theory.  The most important new development on this front is the emergence of numerical plasma codes that self-consistently include radiation reaction effects on the plasma and, simultaneously, compute various observable radiative signatures.  One of the most prominent examples of this is the radiative relativistic PIC code Zeltron developed at the University of Colorado \citep{Cerutti_etal-2013, Cerutti_etal-2014a}.  
In addition, active efforts are now underway to augment various fluid-level (e.g., resistive MHD and two-fluid) codes with radiative modules \citep{Takahashi_Ohsuga-2013, Leake_etal-2013, Sadowski_etal-2014, McKinney_etal-2014, Takahashi_Ohsuga-2015, Ni_etal-2015}, which will enable one to study collisional and optically-thick reconnection problems. 
Importantly, while Zeltron has been developed specifically to study radiative magnetic reconnection, an area in which it has already made important contributions, this code --- and, one hopes, other radiative plasma codes that are being developed or will be developed in the near future --- is sufficiently versatile and can be employed to study other important problems in radiative plasma physics and astrophysics, such as collisionless shocks and turbulence.  
In this sense, our research efforts towards better understanding astrophysical radiative reconnection should not only lead to progress in this particular area, but also should benefit the broader fields of plasma physics and plasma astrophysics. 

Although a lot of progress in developing new radiative computational capabilities has already been achieved, still more work needs to be done. 
Among the most important radiation processes that should, and hopefully will, be incorporated into kinetic plasma codes (in addition to synchrotron and inverse-Compton radiation already implemented in Zeltron) are non-relativistic cyclotron radiation, Klein-Nishina effects for Compton scattering, curvature radiation, and the quantum-electrodynamic modifications to various radiation processes in the presence of a magnetar-strength (above $B_Q$) magnetic field. In addition, there is also strong astrophysical motivation to include collisional and finite optical depth effects such as bremsstrahlung emission and absorption, synchrotron self-absorption, synchrotron-self-Compton (SSC) radiation, and pair creation.  
All these capabilities will greatly expand our ability to study magnetic reconnection, as well as other important plasma processes, in various high-energy astrophysical contexts and thus ultimately will help us attain a better understanding of this violent, shining, beautiful Universe.


\begin{acknowledgements} 
I am very grateful to the organizers of the Parker Workshop on Magnetic Reconnection in Brazil, March 2014, and especially to Dr. Walter Gonzalez.  
I am also indebted to Prof. Eugene Parker for being a constant shining inspiration. 

I am also grateful to numerous colleagues for many stimulating and insightful conversations over many years on various topics discussed in this Chapter.  Specifically, I would like to thank M. Begelman, A. Beloborodov, A. Bhattacharjee,  B. Cerutti, W. Daughton, E. de Gouveia dal Pino, J. Drake, D. Giannios, J. Goodman, R. Kulsrud, H. Li, N. Loureiro, Yu. Lyubarsky, M. Lyutikov, J. McKinney, M. Medvedev, K. Nalewajko,  A. Spitkovsky, and G. Werner.

This work has been supported by NSF Grants PHY-0903851 and AST-1411879	, DOE Grants DE-SC0008409 and DE-SC0008655, and NASA Grants NNX11AE12G,  NNX12AP17G, NNX12AP18G, and NNX13AO83G. 

\end{acknowledgements} 

\bibliographystyle{apj}
\bibliography{rad_recn}

\begin{thebibliography}{159}
\expandafter\ifx\csname natexlab\endcsname\relax\def\natexlab#1{#1}\fi

\bibitem[{{Abdo} {et~al.}(2011){Abdo}, {Ackermann}, {Ajello}, {Allafort},
  {Baldini}, {Ballet}, {Barbiellini}, {Bastieri}, {Bechtol}, {Bellazzini},
  {Berenji}, {Blandford}, {Bloom}, {Bonamente}, {Borgland}, {Bouvier},
  {Brandt}, {Bregeon}, {Brez}, {Brigida}, {Bruel}, {Buehler}, {Buson},
  {Caliandro}, {Cameron}, {Cannon}, {Caraveo}, {Casandjian}, {{\c C}elik},
  {Charles}, {Chekhtman}, {Cheung}, {Chiang}, {Ciprini}, {Claus},
  {Cohen-Tanugi}, {Costamante}, {Cutini}, {D'Ammando}, {Dermer}, {de Angelis},
  {de Luca}, {de Palma}, {Digel}, {do Couto e Silva}, {Drell}, {Drlica-Wagner},
  {Dubois}, {Dumora}, {Favuzzi}, {Fegan}, {Ferrara}, {Focke}, {Fortin},
  {Frailis}, {Fukazawa}, {Funk}, {Fusco}, {Gargano}, {Gasparrini}, {Gehrels},
  {Germani}, {Giglietto}, {Giordano}, {Giroletti}, {Glanzman}, {Godfrey},
  {Grenier}, {Grondin}, {Grove}, {Guiriec}, {Hadasch}, {Hanabata}, {Harding},
  {Hayashi}, {Hayashida}, {Hays}, {Horan}, {Itoh}, {J{\'o}hannesson},
  {Johnson}, {Johnson}, {Khangulyan}, {Kamae}, {Katagiri}, {Kataoka}, {Kerr},
  {Kn{\"o}dlseder}, {Kuss}, {Lande}, {Latronico}, {Lee}, {Lemoine-Goumard},
  {Longo}, {Loparco}, {Lubrano}, {Madejski}, {Makeev}, {Marelli}, {Mazziotta},
  {McEnery}, {Michelson}, {Mitthumsiri}, {Mizuno}, {Moiseev}, {Monte},
  {Monzani}, {Morselli}, {Moskalenko}, {Murgia}, {Nakamori}, {Naumann-Godo},
  {Nolan}, {Norris}, {Nuss}, {Ohsugi}, {Okumura}, {Omodei}, {Ormes}, {Ozaki},
  {Paneque}, {Parent}, {Pelassa}, {Pepe}, {Pesce-Rollins}, {Pierbattista},
  {Piron}, {Porter}, {Rain{\`o}}, {Rando}, {Ray}, {Razzano}, {Reimer},
  {Reimer}, {Reposeur}, {Ritz}, {Romani}, {Sadrozinski}, {Sanchez},
  {Parkinson}, {Scargle}, {Schalk}, {Sgr{\`o}}, {Siskind}, {Smith}, {Spandre},
  {Spinelli}, {Strickman}, {Suson}, {Takahashi}, {Takahashi}, {Tanaka},
  {Thayer}, {Thompson}, {Tibaldo}, {Torres}, {Tosti}, {Tramacere}, {Troja},
  {Uchiyama}, {Vandenbroucke}, {Vasileiou}, {Vianello}, {Vitale}, {Wang},
  {Wood}, {Yang}, \& {Ziegler}}]{Abdo_etal-2011}
{Abdo}, A.~A., {Ackermann}, M., {Ajello}, M., {Allafort}, A., {Baldini}, L.,
  {Ballet}, J., {Barbiellini}, G., {Bastieri}, D., {Bechtol}, K., {Bellazzini},
  R., {Berenji}, B., {Blandford}, R.~D., {Bloom}, E.~D., {Bonamente}, E.,
  {Borgland}, A.~W., {Bouvier}, A., {Brandt}, T.~J., {Bregeon}, J., {Brez}, A.,
  {Brigida}, M., {Bruel}, P., {Buehler}, R., {Buson}, S., {Caliandro}, G.~A.,
  {Cameron}, R.~A., {Cannon}, A., {Caraveo}, P.~A., {Casandjian}, J.~M., {{\c
  C}elik}, {\"O}., {Charles}, E., {Chekhtman}, A., {Cheung}, C.~C., {Chiang},
  J., {Ciprini}, S., {Claus}, R., {Cohen-Tanugi}, J., {Costamante}, L.,
  {Cutini}, S., {D'Ammando}, F., {Dermer}, C.~D., {de Angelis}, A., {de Luca},
  A., {de Palma}, F., {Digel}, S.~W., {do Couto e Silva}, E., {Drell}, P.~S.,
  {Drlica-Wagner}, A., {Dubois}, R., {Dumora}, D., {Favuzzi}, C., {Fegan},
  S.~J., {Ferrara}, E.~C., {Focke}, W.~B., {Fortin}, P., {Frailis}, M.,
  {Fukazawa}, Y., {Funk}, S., {Fusco}, P., {Gargano}, F., {Gasparrini}, D.,
  {Gehrels}, N., {Germani}, S., {Giglietto}, N., {Giordano}, F., {Giroletti},
  M., {Glanzman}, T., {Godfrey}, G., {Grenier}, I.~A., {Grondin}, M.-H.,
  {Grove}, J.~E., {Guiriec}, S., {Hadasch}, D., {Hanabata}, Y., {Harding},
  A.~K., {Hayashi}, K., {Hayashida}, M., {Hays}, E., {Horan}, D., {Itoh}, R.,
  {J{\'o}hannesson}, G., {Johnson}, A.~S., {Johnson}, T.~J., {Khangulyan}, D.,
  {Kamae}, T., {Katagiri}, H., {Kataoka}, J., {Kerr}, M., {Kn{\"o}dlseder}, J.,
  {Kuss}, M., {Lande}, J., {Latronico}, L., {Lee}, S.-H., {Lemoine-Goumard},
  M., {Longo}, F., {Loparco}, F., {Lubrano}, P., {Madejski}, G.~M., {Makeev},
  A., {Marelli}, M., {Mazziotta}, M.~N., {McEnery}, J.~E., {Michelson}, P.~F.,
  {Mitthumsiri}, W., {Mizuno}, T., {Moiseev}, A.~A., {Monte}, C., {Monzani},
  M.~E., {Morselli}, A., {Moskalenko}, I.~V., {Murgia}, S., {Nakamori}, T.,
  {Naumann-Godo}, M., {Nolan}, P.~L., {Norris}, J.~P., {Nuss}, E., {Ohsugi},
  T., {Okumura}, A., {Omodei}, N., {Ormes}, J.~F., {Ozaki}, M., {Paneque}, D.,
  {Parent}, D., {Pelassa}, V., {Pepe}, M., {Pesce-Rollins}, M., {Pierbattista},
  M., {Piron}, F., {Porter}, T.~A., {Rain{\`o}}, S., {Rando}, R., {Ray}, P.~S.,
  {Razzano}, M., {Reimer}, A., {Reimer}, O., {Reposeur}, T., {Ritz}, S.,
  {Romani}, R.~W., {Sadrozinski}, H.~F.-W., {Sanchez}, D., {Parkinson},
  P.~M.~S., {Scargle}, J.~D., {Schalk}, T.~L., {Sgr{\`o}}, C., {Siskind},
  E.~J., {Smith}, P.~D., {Spandre}, G., {Spinelli}, P., {Strickman}, M.~S.,
  {Suson}, D.~J., {Takahashi}, H., {Takahashi}, T., {Tanaka}, T., {Thayer},
  J.~B., {Thompson}, D.~J., {Tibaldo}, L., {Torres}, D.~F., {Tosti}, G.,
  {Tramacere}, A., {Troja}, E., {Uchiyama}, Y., {Vandenbroucke}, J.,
  {Vasileiou}, V., {Vianello}, G., {Vitale}, V., {Wang}, P., {Wood}, K.~S.,
  {Yang}, Z., \& {Ziegler}, M. 2011, Science, 331, 739

\bibitem[{{Aharonian} {et~al.}(2002){Aharonian}, {Belyanin}, {Derishev},
  {Kocharovsky}, \& {Kocharovsky}}]{Aharonian_etal-2002}
{Aharonian}, F.~A., {Belyanin}, A.~A., {Derishev}, E.~V., {Kocharovsky}, V.~V.,
  \& {Kocharovsky}, V.~V. 2002, Phys. Rev. D, 66, 023005

\bibitem[{{Arka} \& {Dubus}(2013)}]{Arka_Dubus-2013}
{Arka}, I. \& {Dubus}, G. 2013, Astron. \& Astrophys., 550, A101

\bibitem[{{Axford}(1967)}]{Axford-1967}
{Axford}, W.~I. 1967, Space Sci. Rev., 7, 149

\bibitem[{{Balbo} {et~al.}(2011){Balbo}, {Walter}, {Ferrigno}, \&
  {Bordas}}]{Balbo_etal-2011}
{Balbo}, M., {Walter}, R., {Ferrigno}, C., \& {Bordas}, P. 2011, Astron. \&
  Astrophys., 527, L4

\bibitem[{{Bednarek} \& {Idec}(2011)}]{Bednarek_Idec-2011}
{Bednarek}, W. \& {Idec}, W. 2011, Mon. Not. Roy. Astron. Soc., 414, 2229

\bibitem[{{Bessho} \& {Bhattacharjee}(2012)}]{Bessho_Bhattacharjee-2012}
{Bessho}, N. \& {Bhattacharjee}, A. 2012, Astrophys. J., 750, 129

\bibitem[{{Bhattacharjee} {et~al.}(2009){Bhattacharjee}, {Huang}, {Yang}, \&
  {Rogers}}]{Bhattacharjee_etal-2009}
{Bhattacharjee}, A., {Huang}, Y., {Yang}, H., \& {Rogers}, B. 2009, Phys.
  Plasmas, 16, 112102

\bibitem[{{Birn} {et~al.}(2001){Birn}, {Drake}, {Shay}, {Rogers}, {Denton},
  {Hesse}, {Kuznetsova}, {Ma}, {Bhattacharjee}, {Otto}, \&
  {Pritchett}}]{Birn_etal-2001}
{Birn}, J., {Drake}, J.~F., {Shay}, M.~A., {Rogers}, B.~N., {Denton}, R.~E.,
  {Hesse}, M., {Kuznetsova}, M., {Ma}, Z.~W., {Bhattacharjee}, A., {Otto}, A.,
  \& {Pritchett}, P.~L. 2001, J. Geophys. Res., 106, 3715

\bibitem[{{Biskamp}(2000)}]{Biskamp-2000}
{Biskamp}, D. 2000, {Magnetic Reconnection in Plasmas}

\bibitem[{{Blumenthal} \& {Gould}(1970)}]{Blumenthal_Gould-1970}
{Blumenthal}, G.~R. \& {Gould}, R.~J. 1970, Reviews of Modern Physics, 42, 237

\bibitem[{{Buehler} {et~al.}(2012){Buehler}, {Scargle}, {Blandford}, {Baldini},
  {Baring}, {Belfiore}, {Charles}, {Chiang}, {D'Ammando}, {Dermer}, {Funk},
  {Grove}, {Harding}, {Hays}, {Kerr}, {Massaro}, {Mazziotta}, {Romani}, {Saz
  Parkinson}, {Tennant}, \& {Weisskopf}}]{Buehler_etal-2012}
{Buehler}, R., {Scargle}, J.~D., {Blandford}, R.~D., {Baldini}, L., {Baring},
  M.~G., {Belfiore}, A., {Charles}, E., {Chiang}, J., {D'Ammando}, F.,
  {Dermer}, C.~D., {Funk}, S., {Grove}, J.~E., {Harding}, A.~K., {Hays}, E.,
  {Kerr}, M., {Massaro}, F., {Mazziotta}, M.~N., {Romani}, R.~W., {Saz
  Parkinson}, P.~M., {Tennant}, A.~F., \& {Weisskopf}, M.~C. 2012, Astrophys.
  J., 749, 26

\bibitem[{{B{\"u}hler} \& {Blandford}(2014)}]{Buehler_Blandford-2014}
{B{\"u}hler}, R. \& {Blandford}, R. 2014, Reports on Progress in Physics, 77,
  066901

\bibitem[{{Bykov} {et~al.}(2012){Bykov}, {Pavlov}, {Artemyev}, \&
  {Uvarov}}]{Bykov_etal-2012}
{Bykov}, A.~M., {Pavlov}, G.~G., {Artemyev}, A.~V., \& {Uvarov}, Y.~A. 2012,
  Mon. Not. Roy. Astron. Soc., 421, L67

\bibitem[{{Cerutti} {et~al.}(2015){Cerutti}, {Philippov}, {Parfrey}, \&
  {Spitkovsky}}]{Cerutti_etal-2015}
{Cerutti}, B., {Philippov}, A., {Parfrey}, K., \& {Spitkovsky}, A. 2015, Mon.
  Not. Roy. Astron. Soc., 448, 606

\bibitem[{{Cerutti} {et~al.}(2012{\natexlab{a}}){Cerutti}, {Uzdensky}, \&
  {Begelman}}]{Cerutti_etal-2012a}
{Cerutti}, B., {Uzdensky}, D.~A., \& {Begelman}, M.~C. 2012{\natexlab{a}},
  Astrophys. J., 746, 148

\bibitem[{{Cerutti} {et~al.}(2012{\natexlab{b}}){Cerutti}, {Werner},
  {Uzdensky}, \& {Begelman}}]{Cerutti_etal-2012b}
{Cerutti}, B., {Werner}, G.~R., {Uzdensky}, D.~A., \& {Begelman}, M.~C.
  2012{\natexlab{b}}, Astrophys. J. Lett., 754, L33

\bibitem[{{Cerutti} {et~al.}(2013){Cerutti}, {Werner}, {Uzdensky}, \&
  {Begelman}}]{Cerutti_etal-2013}
---. 2013, Astrophys. J., 770, 147

\bibitem[{{Cerutti} {et~al.}(2014{\natexlab{a}}){Cerutti}, {Werner},
  {Uzdensky}, \& {Begelman}}]{Cerutti_etal-2014b}
---. 2014{\natexlab{a}}, Physics of Plasmas, 21, 056501

\bibitem[{{Cerutti} {et~al.}(2014{\natexlab{b}}){Cerutti}, {Werner},
  {Uzdensky}, \& {Begelman}}]{Cerutti_etal-2014a}
---. 2014{\natexlab{b}}, Astrophys. J., 782, 104

\bibitem[{{Chugunov} {et~al.}(1975){Chugunov}, {Eidman}, \&
  {Suvorov}}]{Chugunov_etal-1975}
{Chugunov}, I.~V., {Eidman}, V.~I., \& {Suvorov}, E.~V. 1975, Astrophys. \&
  Space Sci., 32, L7

\bibitem[{{Clausen-Brown} \& {Lyutikov}(2012)}]{Clausen-Brown_Lyutikov-2012}
{Clausen-Brown}, E. \& {Lyutikov}, M. 2012, Mon. Not. Roy. Astron. Soc., 426,
  1374

\bibitem[{{Contopoulos}(2007)}]{Contopoulos-2007}
{Contopoulos}, I. 2007, Astron. \& Astrophys., 466, 301

\bibitem[{{Coroniti}(1990)}]{Coroniti-1990}
{Coroniti}, F.~V. 1990, Astrophys. J., 349, 538

\bibitem[{{de Gouveia dal Pino} \& {Lazarian}(2005)}]{deGouveia_etal-2005}
{de Gouveia dal Pino}, E.~M. \& {Lazarian}, A. 2005, Astron. \& Astrophys.,
  441, 845

\bibitem[{{de Gouveia Dal Pino} {et~al.}(2010){de Gouveia Dal Pino},
  {Piovezan}, \& {Kadowaki}}]{deGouveia_etal-2010}
{de Gouveia Dal Pino}, E.~M., {Piovezan}, P.~P., \& {Kadowaki}, L.~H.~S. 2010,
  Astron. \& Astrophys., 518, A5

\bibitem[{{de Jager} {et~al.}(1996){de Jager}, {Harding}, {Michelson}, {Nel},
  {Nolan}, {Sreekumar}, \& {Thompson}}]{deJager_etal-1996}
{de Jager}, O.~C., {Harding}, A.~K., {Michelson}, P.~F., {Nel}, H.~I., {Nolan},
  P.~L., {Sreekumar}, P., \& {Thompson}, D.~J. 1996, Astrophys. J., 457, 253

\bibitem[{{Di Matteo}(1998)}]{DiMatteo-1998}
{Di Matteo}, T. 1998, Mon. Not. Roy. Astron. Soc., 299, L15

\bibitem[{{Di Matteo} {et~al.}(1999){Di Matteo}, {Celotti}, \&
  {Fabian}}]{DiMatteo_etal-1999}
{Di Matteo}, T., {Celotti}, A., \& {Fabian}, A.~C. 1999, Mon. Not. Roy. Astron.
  Soc., 304, 809

\bibitem[{{Dong} {et~al.}(2012){Dong}, {Wang}, {Lu}, {Huang}, {Yuan}, {Liu},
  {Lin}, {Li}, {Wei}, {Zhong}, {Shi}, {Jiang}, {Ding}, {Jiang}, {Du}, {He},
  {Yu}, {Liu}, {Wang}, {Tang}, {Zhu}, {Zhao}, {Sheng}, \&
  {Zhang}}]{Dong_etal-2012}
{Dong}, Q.-L., {Wang}, S.-J., {Lu}, Q.-M., {Huang}, C., {Yuan}, D.-W., {Liu},
  X., {Lin}, X.-X., {Li}, Y.-T., {Wei}, H.-G., {Zhong}, J.-Y., {Shi}, J.-R.,
  {Jiang}, S.-E., {Ding}, Y.-K., {Jiang}, B.-B., {Du}, K., {He}, X.-T., {Yu},
  M.~Y., {Liu}, C.~S., {Wang}, S., {Tang}, Y.-J., {Zhu}, J.-Q., {Zhao}, G.,
  {Sheng}, Z.-M., \& {Zhang}, J. 2012, Physical Review Letters, 108, 215001

\bibitem[{{Dorman} \& {Kulsrud}(1995)}]{Dorman_Kulsrud-1995}
{Dorman}, V.~L. \& {Kulsrud}, R.~M. 1995, Astrophys. J., 449, 777

\bibitem[{{Drenkhahn} \& {Spruit}(2002)}]{Drenkhahn_Spruit-2002}
{Drenkhahn}, G. \& {Spruit}, H.~C. 2002, Astron. \& Astrophys., 391, 1141

\bibitem[{{Dungey}(1961)}]{Dungey-1961}
{Dungey}, J.~W. 1961, Phys. Rev. Lett., 6, 47

\bibitem[{{Eyink} {et~al.}(2011){Eyink}, {Lazarian}, \&
  {Vishniac}}]{Eyink_etal-2011}
{Eyink}, G.~L., {Lazarian}, A., \& {Vishniac}, E.~T. 2011, Astrophys. J., 743,
  51

\bibitem[{{Fiksel} {et~al.}(2014){Fiksel}, {Fox}, {Bhattacharjee}, {Barnak},
  {Chang}, {Germaschewski}, {Hu}, \& {Nilson}}]{Fiksel_etal-2014}
{Fiksel}, G., {Fox}, W., {Bhattacharjee}, A., {Barnak}, D.~H., {Chang}, P.-Y.,
  {Germaschewski}, K., {Hu}, S.~X., \& {Nilson}, P.~M. 2014, Physical Review
  Letters, 113, 105003

\bibitem[{{Fox} {et~al.}(2011){Fox}, {Bhattacharjee}, \&
  {Germaschewski}}]{Fox_etal-2011}
{Fox}, W., {Bhattacharjee}, A., \& {Germaschewski}, K. 2011, Physical Review
  Letters, 106, 215003

\bibitem[{{Fox} {et~al.}(2012){Fox}, {Bhattacharjee}, \&
  {Germaschewski}}]{Fox_etal-2012}
---. 2012, Physics of Plasmas, 19, 056309

\bibitem[{{Galeev} {et~al.}(1979){Galeev}, {Rosner}, \& {Vaiana}}]{GRV-1979}
{Galeev}, A.~A., {Rosner}, R., \& {Vaiana}, G.~S. 1979, Astrophys. J., 229, 318

\bibitem[{{Giannios}(2010)}]{Giannios-2010}
{Giannios}, D. 2010, Mon. Not. Roy. Astron. Soc., 408, L46

\bibitem[{{Giannios}(2013)}]{Giannios-2013}
---. 2013, Mon. Not. Roy. Astron. Soc., 431, 355

\bibitem[{{Giannios} \& {Spruit}(2006)}]{Giannios_Spruit-2006}
{Giannios}, D. \& {Spruit}, H.~C. 2006, Astron. \& Astrophys., 450, 887

\bibitem[{{Giannios} \& {Spruit}(2007)}]{Giannios_Spruit-2007}
---. 2007, Astron. \& Astrophys., 469, 1

\bibitem[{{Giannios} {et~al.}(2009){Giannios}, {Uzdensky}, \&
  {Begelman}}]{Giannios_etal-2009}
{Giannios}, D., {Uzdensky}, D.~A., \& {Begelman}, M.~C. 2009, Mon. Not. Roy.
  Astron. Soc., 395, L29

\bibitem[{{Giannios} {et~al.}(2010){Giannios}, {Uzdensky}, \&
  {Begelman}}]{Giannios_etal-2010}
---. 2010, Mon. Not. Roy. Astron. Soc., 402, 1649

\bibitem[{{Goodman} \& {Uzdensky}(2008)}]{Goodman_Uzdensky-2008}
{Goodman}, J. \& {Uzdensky}, D. 2008, Astrophys. J., 688, 555

\bibitem[{{Guilbert} {et~al.}(1983){Guilbert}, {Fabian}, \&
  {Rees}}]{Guilbert_etal-1983}
{Guilbert}, P.~W., {Fabian}, A.~C., \& {Rees}, M.~J. 1983, Mon. Not. Roy.
  Astron. Soc., 205, 593

\bibitem[{{Guo} {et~al.}(2014){Guo}, {Li}, {Daughton}, \&
  {Liu}}]{Guo_etal-2014}
{Guo}, F., {Li}, H., {Daughton}, W., \& {Liu}, Y.-H. 2014, Physical Review
  Letters, 113, 155005

\bibitem[{{Hillas}(1984)}]{Hillas-1984}
{Hillas}, A.~M. 1984, Annual Rev. Astron. \& Astrophys., 22, 425

\bibitem[{{Hoshino} \& {Lyubarsky}(2012)}]{Hoshino_Lyubarsky-2012}
{Hoshino}, M. \& {Lyubarsky}, Y. 2012, Space Sci. Rev., 173, 521

\bibitem[{{Huang} \& {Bhattacharjee}(2010)}]{Huang_Bhattacharjee-2010}
{Huang}, Y. \& {Bhattacharjee}, A. 2010, Physics of Plasmas, 17, 062104

\bibitem[{{Jackson}(1975)}]{Jackson-1975}
{Jackson}, J.~D. 1975, {Classical electrodynamics}

\bibitem[{{Jaroschek} \& {Hoshino}(2009)}]{Jaroschek_Hoshino-2009}
{Jaroschek}, C.~H. \& {Hoshino}, M. 2009, Physical Review Letters, 103, 075002

\bibitem[{{Jaroschek} {et~al.}(2004{\natexlab{a}}){Jaroschek}, {Lesch}, \&
  {Treumann}}]{Jaroschek_etal-2004b}
{Jaroschek}, C.~H., {Lesch}, H., \& {Treumann}, R.~A. 2004{\natexlab{a}},
  Astrophys. J. Lett., 605, L9

\bibitem[{{Jaroschek} {et~al.}(2004{\natexlab{b}}){Jaroschek}, {Treumann},
  {Lesch}, \& {Scholer}}]{Jaroschek_etal-2004a}
{Jaroschek}, C.~H., {Treumann}, R.~A., {Lesch}, H., \& {Scholer}, M.
  2004{\natexlab{b}}, Phys. Plasmas, 11, 1151

\bibitem[{{Kadowaki} {et~al.}(2015){Kadowaki}, {de Gouveia Dal Pino}, \&
  {Singh}}]{Kadowaki_etal-2015}
{Kadowaki}, L.~H.~S., {de Gouveia Dal Pino}, E.~M., \& {Singh}, C.~B. 2015,
  Astrophys. J., 802, 113

\bibitem[{{Kagan} {et~al.}(2013){Kagan}, {Milosavljevi{\'c}}, \&
  {Spitkovsky}}]{Kagan_etal-2013}
{Kagan}, D., {Milosavljevi{\'c}}, M., \& {Spitkovsky}, A. 2013, Astrophys. J.,
  774, 41

\bibitem[{{Khiali} {et~al.}(2015{\natexlab{a}}){Khiali}, {de Gouveia Dal Pino},
  \& {del Valle}}]{Khiali_etal-2015a}
{Khiali}, B., {de Gouveia Dal Pino}, E.~M., \& {del Valle}, M.~V.
  2015{\natexlab{a}}, Mon. Not. Roy. Astron. Soc., 449, 34

\bibitem[{{Khiali} {et~al.}(2015{\natexlab{b}}){Khiali}, {de Gouveia Dal Pino},
  \& {Sol}}]{Khiali_etal-2015b}
{Khiali}, B., {de Gouveia Dal Pino}, E.~M., \& {Sol}, H. 2015{\natexlab{b}},
  ArXiv e-prints

\bibitem[{{Kirk}(2004)}]{Kirk-2004}
{Kirk}, J.~G. 2004, Physical Review Letters, 92, 181101

\bibitem[{{Kirk} \& {Skj{\ae}raasen}(2003)}]{Kirk_Skjaeraasen-2003}
{Kirk}, J.~G. \& {Skj{\ae}raasen}, O. 2003, Astrophys. J., 591, 366

\bibitem[{{Komissarov} \& {Lyutikov}(2011)}]{Komissarov_Lyutikov-2011}
{Komissarov}, S.~S. \& {Lyutikov}, M. 2011, Mon. Not. Roy. Astron. Soc., 414,
  2017

\bibitem[{{Kowal} {et~al.}(2009){Kowal}, {Lazarian}, {Vishniac}, \&
  {Otmianowska-Mazur}}]{Kowal_etal-2009}
{Kowal}, G., {Lazarian}, A., {Vishniac}, E.~T., \& {Otmianowska-Mazur}, K.
  2009, Astrophys. J., 700, 63

\bibitem[{{Krucker} {et~al.}(2010){Krucker}, {Hudson}, {Glesener}, {White},
  {Masuda}, {Wuelser}, \& {Lin}}]{Krucker_etal-2010}
{Krucker}, S., {Hudson}, H.~S., {Glesener}, L., {White}, S.~M., {Masuda}, S.,
  {Wuelser}, J.-P., \& {Lin}, R.~P. 2010, Astrophys. J., 714, 1108

\bibitem[{{Landau} \& {Lifshitz}(1971)}]{Landau_Lifshitz-1971}
{Landau}, L.~D. \& {Lifshitz}, E.~M. 1971, {The classical theory of fields}

\bibitem[{{Larrabee} {et~al.}(2003){Larrabee}, {Lovelace}, \&
  {Romanova}}]{Larrabee_etal-2003}
{Larrabee}, D.~A., {Lovelace}, R.~V.~E., \& {Romanova}, M.~M. 2003, Astrophys.
  J., 586, 72

\bibitem[{{Lazarian} \& {Vishniac}(1999)}]{Lazarian_Vishniac-1999}
{Lazarian}, A. \& {Vishniac}, E.~T. 1999, Astrophys. J., 517, 700

\bibitem[{{Leake} {et~al.}(2013){Leake}, {Lukin}, \&
  {Linton}}]{Leake_etal-2013}
{Leake}, J.~E., {Lukin}, V.~S., \& {Linton}, M.~G. 2013, Physics of Plasmas,
  20, 061202

\bibitem[{{Lebedev} {et~al.}(2014){Lebedev}, {Suzuki-Vidal}, {Pickworth},
  {Swadling}, {Burdiak}, {Skidmore}, {Hall}, {Bennett}, {Bland}, {Chittenden},
  {de Grouchy}, {Derrick}, {Hare}, {Parker}, {Sciortino}, {Suttle}, {Ciardi},
  {Rodriguez}, {Gil}, {Espinosa}, {Hansen}, {Frank}, \&
  {Music}}]{Lebedev_etal-2014}
{Lebedev}, S.~V., {Suzuki-Vidal}, F., {Pickworth}, L.~A., {Swadling}, G.~F.,
  {Burdiak}, G., {Skidmore}, J., {Hall}, G.~N., {Bennett}, M., {Bland}, S.~N.,
  {Chittenden}, J.~P., {de Grouchy}, P., {Derrick}, J., {Hare}, J., {Parker},
  T., {Sciortino}, F., {Suttle}, L., {Ciardi}, A., {Rodriguez}, R., {Gil},
  J.~M., {Espinosa}, G., {Hansen}, E., {Frank}, A., \& {Music}, J. 2014, in APS
  Meeting Abstracts, 8107P

\bibitem[{{Lesch} \& {Birk}(1998)}]{Lesch_Birk-1998}
{Lesch}, H. \& {Birk}, G.~T. 1998, Astrophys. J., 499, 167

\bibitem[{{Li} {et~al.}(2007){Li}, {S{\'e}guin}, {Frenje}, {Rygg}, {Petrasso},
  {Town}, {Landen}, {Knauer}, \& {Smalyuk}}]{Li_etal-2007}
{Li}, C.~K., {S{\'e}guin}, F.~H., {Frenje}, J.~A., {Rygg}, J.~R., {Petrasso},
  R.~D., {Town}, R.~P.~J., {Landen}, O.~L., {Knauer}, J.~P., \& {Smalyuk},
  V.~A. 2007, Physical Review Letters, 99, 055001

\bibitem[{{Liu} {et~al.}(2003){Liu}, {Mineshige}, \& {Ohsuga}}]{Liu_etal-2003}
{Liu}, B.~F., {Mineshige}, S., \& {Ohsuga}, K. 2003, Astrophys. J., 587, 571

\bibitem[{{Liu} {et~al.}(2011){Liu}, {Li}, {Yin}, {Albright}, {Bowers}, \&
  {Liang}}]{Liu_etal-2011}
{Liu}, W., {Li}, H., {Yin}, L., {Albright}, B.~J., {Bowers}, K.~J., \& {Liang},
  E.~P. 2011, Physics of Plasmas, 18, 052105

\bibitem[{{Loureiro} {et~al.}(2012){Loureiro}, {Samtaney}, {Schekochihin}, \&
  {Uzdensky}}]{Loureiro_etal-2012}
{Loureiro}, N.~F., {Samtaney}, R., {Schekochihin}, A.~A., \& {Uzdensky}, D.~A.
  2012, Physics of Plasmas, 19, 042303

\bibitem[{{Loureiro} {et~al.}(2007){Loureiro}, {Schekochihin}, \&
  {Cowley}}]{Loureiro_etal-2007}
{Loureiro}, N.~F., {Schekochihin}, A.~A., \& {Cowley}, S.~C. 2007, Phys.
  Plasmas, 14, 100703

\bibitem[{{Loureiro} {et~al.}(2009){Loureiro}, {Uzdensky}, {Schekochihin},
  {Cowley}, \& {Yousef}}]{Loureiro_etal-2009}
{Loureiro}, N.~F., {Uzdensky}, D.~A., {Schekochihin}, A.~A., {Cowley}, S.~C.,
  \& {Yousef}, T.~A. 2009, Mon. Not. Roy. Astron. Soc. Lett., 399, L146

\bibitem[{{Lyubarskii}(1996)}]{Lyubarsky-1996}
{Lyubarskii}, Y.~E. 1996, Astron. \& Astrophys., 311, 172

\bibitem[{{Lyubarskii}(2000)}]{Lyubarsky-2000}
{Lyubarskii}, Y.~E. 2000, in Astronomical Society of the Pacific Conference
  Series, Vol. 202, IAU Colloq. 177: Pulsar Astronomy - 2000 and Beyond, ed.
  M.~{Kramer}, N.~{Wex}, \& R.~{Wielebinski}, 439

\bibitem[{{Lyubarsky} \& {Kirk}(2001)}]{Lyubarsky_Kirk-2001}
{Lyubarsky}, Y. \& {Kirk}, J.~G. 2001, Astrophys. J., 547, 437

\bibitem[{{Lyubarsky} \& {Liverts}(2008)}]{Lyubarsky_Liverts-2008}
{Lyubarsky}, Y. \& {Liverts}, M. 2008, Astrophys. J., 682, 1436

\bibitem[{{Lyubarsky}(2012)}]{Lyubarsky-2012}
{Lyubarsky}, Y.~E. 2012, Mon. Not. Roy. Astron. Soc., 427, 1497

\bibitem[{{Lyutikov}(2003{\natexlab{a}})}]{Lyutikov-2003b}
{Lyutikov}, M. 2003{\natexlab{a}}, Mon. Not. Roy. Astron. Soc., 346, 540

\bibitem[{{Lyutikov}(2003{\natexlab{b}})}]{Lyutikov-2003a}
---. 2003{\natexlab{b}}, New Astron. Rev, 47, 513

\bibitem[{{Lyutikov}(2006{\natexlab{a}})}]{Lyutikov-2006a}
---. 2006{\natexlab{a}}, Mon. Not. Roy. Astron. Soc., 367, 1594

\bibitem[{{Lyutikov}(2006{\natexlab{b}})}]{Lyutikov-2006b}
---. 2006{\natexlab{b}}, New Journal of Physics, 8, 119

\bibitem[{{Lyutikov}(2010)}]{Lyutikov-2010}
---. 2010, Mon. Not. Roy. Astron. Soc., 405, 1809

\bibitem[{{Lyutikov} {et~al.}(2012{\natexlab{a}}){Lyutikov}, {Balsara}, \&
  {Matthews}}]{Lyutikov_etal-2012a}
{Lyutikov}, M., {Balsara}, D., \& {Matthews}, C. 2012{\natexlab{a}}, Mon. Not.
  Roy. Astron. Soc., 422, 3118

\bibitem[{{Lyutikov} {et~al.}(2012{\natexlab{b}}){Lyutikov}, {Otte}, \&
  {McCann}}]{Lyutikov_etal-2012b}
{Lyutikov}, M., {Otte}, N., \& {McCann}, A. 2012{\natexlab{b}}, Astrophys. J.,
  754, 33

\bibitem[{{Masada} {et~al.}(2010){Masada}, {Nagataki}, {Shibata}, \&
  {Terasawa}}]{Masada_etal-2010}
{Masada}, Y., {Nagataki}, S., {Shibata}, K., \& {Terasawa}, T. 2010, Pub.
  Astron. Soc. Japan, 62, 1093

\bibitem[{{Masuda} {et~al.}(1994){Masuda}, {Kosugi}, {Hara}, {Tsuneta}, \&
  {Ogawara}}]{Masuda_etal-1994}
{Masuda}, S., {Kosugi}, T., {Hara}, H., {Tsuneta}, S., \& {Ogawara}, Y. 1994,
  Nature, 371, 495

\bibitem[{{McKinney} {et~al.}(2014){McKinney}, {Tchekhovskoy}, {Sadowski}, \&
  {Narayan}}]{McKinney_etal-2014}
{McKinney}, J.~C., {Tchekhovskoy}, A., {Sadowski}, A., \& {Narayan}, R. 2014,
  Mon. Not. Roy. Astron. Soc., 441, 3177

\bibitem[{{McKinney} \& {Uzdensky}(2012)}]{McKinney_Uzdensky-2012}
{McKinney}, J.~C. \& {Uzdensky}, D.~A. 2012, Mon. Not. Roy. Astron. Soc., 419,
  573

\bibitem[{{Medvedev}(2000)}]{Medvedev-2000}
{Medvedev}, M.~V. 2000, Astrophys. J., 540, 704

\bibitem[{{Melzani} {et~al.}(2014){Melzani}, {Walder}, {Folini},
  {Winisdoerffer}, \& {Favre}}]{Melzani_etal-2014b}
{Melzani}, M., {Walder}, R., {Folini}, D., {Winisdoerffer}, C., \& {Favre},
  J.~M. 2014, Astron. \& Astrophys., 570, A112

\bibitem[{{Michel}(1982)}]{Michel-1982}
{Michel}, F.~C. 1982, Reviews of Modern Physics, 54, 1

\bibitem[{{Michel}(1994)}]{Michel-1994}
---. 1994, Astrophys. J., 431, 397

\bibitem[{{Nalewajko} {et~al.}(2011){Nalewajko}, {Giannios}, {Begelman},
  {Uzdensky}, \& {Sikora}}]{Nalewajko_etal-2011}
{Nalewajko}, K., {Giannios}, D., {Begelman}, M.~C., {Uzdensky}, D.~A., \&
  {Sikora}, M. 2011, Mon. Not. Roy. Astron. Soc., 413, 333

\bibitem[{{Nalewajko} {et~al.}(2015){Nalewajko}, {Uzdensky}, {Cerutti},
  {Werner}, \& {Begelman}}]{Nalewajko_etal-2015}
{Nalewajko}, K., {Uzdensky}, D.~A., {Cerutti}, B., {Werner}, G.~R., \&
  {Begelman}, M.~C. 2015, ArXiv e-prints

\bibitem[{{Ni} {et~al.}(2015){Ni}, {Kliem}, {Lin}, \& {Wu}}]{Ni_etal-2015}
{Ni}, L., {Kliem}, B., {Lin}, J., \& {Wu}, N. 2015, Astrophys. J., 799, 79

\bibitem[{{Nilson} {et~al.}(2006){Nilson}, {Willingale}, {Kaluza},
  {Kamperidis}, {Minardi}, {Wei}, {Fernandes}, {Notley}, {Bandyopadhyay},
  {Sherlock}, {Kingham}, {Tatarakis}, {Najmudin}, {Rozmus}, {Evans}, {Haines},
  {Dangor}, \& {Krushelnick}}]{Nilson_etal-2006}
{Nilson}, P.~M., {Willingale}, L., {Kaluza}, M.~C., {Kamperidis}, C.,
  {Minardi}, S., {Wei}, M.~S., {Fernandes}, P., {Notley}, M., {Bandyopadhyay},
  S., {Sherlock}, M., {Kingham}, R.~J., {Tatarakis}, M., {Najmudin}, Z.,
  {Rozmus}, W., {Evans}, R.~G., {Haines}, M.~G., {Dangor}, A.~E., \&
  {Krushelnick}, K. 2006, Physical Review Letters, 97, 255001

\bibitem[{{Nilson} {et~al.}(2008){Nilson}, {Willingale}, {Kaluza},
  {Kamperidis}, {Minardi}, {Wei}, {Fernandes}, {Notley}, {Bandyopadhyay},
  {Sherlock}, {Kingham}, {Tatarakis}, {Najmudin}, {Rozmus}, {Evans}, {Haines},
  {Dangor}, \& {Krushelnick}}]{Nilson_etal-2008}
---. 2008, Physics of Plasmas, 15, 092701

\bibitem[{{Oka} {et~al.}(2013){Oka}, {Ishikawa}, {Saint-Hilaire}, {Krucker}, \&
  {Lin}}]{Oka_etal-2013}
{Oka}, M., {Ishikawa}, S., {Saint-Hilaire}, P., {Krucker}, S., \& {Lin}, R.~P.
  2013, Astrophys. J., 764, 6

\bibitem[{{Oka} {et~al.}(2015){Oka}, {Krucker}, {Hudson}, \&
  {Saint-Hilaire}}]{Oka_etal-2015}
{Oka}, M., {Krucker}, S., {Hudson}, H.~S., \& {Saint-Hilaire}, P. 2015,
  Astrophys. J., 799, 129

\bibitem[{{Parfrey} {et~al.}(2012){Parfrey}, {Beloborodov}, \&
  {Hui}}]{Parfrey_etal-2012}
{Parfrey}, K., {Beloborodov}, A.~M., \& {Hui}, L. 2012, Astrophys. J. Lett.,
  754, L12

\bibitem[{{Parfrey} {et~al.}(2013){Parfrey}, {Beloborodov}, \&
  {Hui}}]{Parfrey_etal-2013}
---. 2013, Astrophys. J., 774, 92

\bibitem[{{Parker}(1957)}]{Parker-1957}
{Parker}, E.~N. 1957, J. Geophys. Res., 62, 509

\bibitem[{{Paschmann} {et~al.}(2013){Paschmann}, {{\O}ieroset}, \&
  {Phan}}]{Paschmann_etal-2013}
{Paschmann}, G., {{\O}ieroset}, M., \& {Phan}, T. 2013, Space Sci. Rev., 178,
  385

\bibitem[{{P{\'e}tri} \& {Lyubarsky}(2007)}]{Petri_Lyubarsky-2007}
{P{\'e}tri}, J. \& {Lyubarsky}, Y. 2007, Astron. \& Astrophys., 473, 683

\bibitem[{{Philippov} \& {Spitkovsky}(2014)}]{Philippov_etal-2014}
{Philippov}, A.~A. \& {Spitkovsky}, A. 2014, Astrophys. J. Lett., 785, L33

\bibitem[{{Philippov} {et~al.}(2015){Philippov}, {Spitkovsky}, \&
  {Cerutti}}]{Philippov_etal-2015}
{Philippov}, A.~A., {Spitkovsky}, A., \& {Cerutti}, B. 2015, Astrophys. J.
  Lett., 801, L19

\bibitem[{{Pozdnyakov} {et~al.}(1983){Pozdnyakov}, {Sobol}, \&
  {Syunyaev}}]{Pozdnyakov_etal-1983}
{Pozdnyakov}, L.~A., {Sobol}, I.~M., \& {Syunyaev}, R.~A. 1983, Astrophysics
  and Space Physics Reviews, 2, 189

\bibitem[{{Remillard} \& {McClintock}(2006)}]{Remillard_McClintock-2006}
{Remillard}, R.~A. \& {McClintock}, J.~E. 2006, Annual Rev. Astron. \&
  Astrophys., 44, 49

\bibitem[{{Romanova} \& {Lovelace}(1992)}]{Romanova_Lovelace-1992}
{Romanova}, M.~M. \& {Lovelace}, R.~V.~E. 1992, Astron. \& Astrophys., 262, 26

\bibitem[{{Rybicki} \& {Lightman}(1979)}]{RL-1979}
{Rybicki}, G.~B. \& {Lightman}, A.~P. 1979, {Radiative Processes in
  Astrophysics} (Wiley \& Sons: New York)

\bibitem[{{Sadowski} {et~al.}(2014){Sadowski}, {Narayan}, {McKinney}, \&
  {Tchekhovskoy}}]{Sadowski_etal-2014}
{Sadowski}, A., {Narayan}, R., {McKinney}, J.~C., \& {Tchekhovskoy}, A. 2014,
  Mon. Not. Roy. Astron. Soc., 439, 503

\bibitem[{{Schopper} {et~al.}(1998){Schopper}, {Lesch}, \&
  {Birk}}]{Schopper_etal-1998}
{Schopper}, R., {Lesch}, H., \& {Birk}, G.~T. 1998, Astron. \& Astrophys., 335,
  26

\bibitem[{{Shakura} \& {Sunyaev}(1973)}]{Shakura_Sunyaev-1973}
{Shakura}, N.~I. \& {Sunyaev}, R.~A. 1973, Astron. \& Astrophys., 24, 337

\bibitem[{{Shibata}(1996)}]{Shibata-1996}
{Shibata}, K. 1996, Advances in Space Research, 17, 9

\bibitem[{{Shklovskii}(1957)}]{Shklovsky-1957}
{Shklovskii}, I.~S. 1957, Sov. Astron., 1, 690

\bibitem[{{Shklovskii}(1960)}]{Shklovsky-1960}
---. 1960, {Cosmic radio waves}

\bibitem[{{Shklovskii}(1966)}]{Shklovsky-1966}
---. 1966, Sov. Astron., 10, 6

\bibitem[{{Shklovsky}(1947)}]{Shklovsky-1947}
{Shklovsky}, J.~S. 1947, Nature, 159, 752

\bibitem[{{Singh} {et~al.}(2015){Singh}, {de Gouveia Dal Pino}, \&
  {Kadowaki}}]{Singh_etal-2015}
{Singh}, C.~B., {de Gouveia Dal Pino}, E.~M., \& {Kadowaki}, L.~H.~S. 2015,
  Astrophys. J. Lett., 799, L20

\bibitem[{{Sironi} \& {Spitkovsky}(2011)}]{Sironi_Spitkovsky-2011}
{Sironi}, L. \& {Spitkovsky}, A. 2011, Astrophys. J., 741, 39

\bibitem[{{Sironi} \& {Spitkovsky}(2014)}]{Sironi_Spitkovsky-2014}
---. 2014, Astrophys. J. Lett., 783, L21

\bibitem[{{Spruit} {et~al.}(2001){Spruit}, {Daigne}, \&
  {Drenkhahn}}]{Spruit_etal-2001}
{Spruit}, H.~C., {Daigne}, F., \& {Drenkhahn}, G. 2001, Astron. \& Astrophys.,
  369, 694

\bibitem[{{Steinolfson} \& {van Hoven}(1984)}]{Steinolfson_vanHoven-1984}
{Steinolfson}, R.~S. \& {van Hoven}, G. 1984, Astrophys. J., 276, 391

\bibitem[{{Striani} {et~al.}(2011){Striani}, {Tavani}, {Piano}, {Donnarumma},
  {Pucella}, {Vittorini}, {Bulgarelli}, {Trois}, {Pittori}, {Verrecchia},
  {Costa}, {Weisskopf}, {Tennant}, {Argan}, {Barbiellini}, {Caraveo},
  {Cardillo}, {Cattaneo}, {Chen}, {De Paris}, {Del Monte}, {Di Cocco},
  {Evangelista}, {Ferrari}, {Feroci}, {Fuschino}, {Galli}, {Gianotti},
  {Giuliani}, {Labanti}, {Lapshov}, {Lazzarotto}, {Longo}, {Marisaldi},
  {Mereghetti}, {Morselli}, {Pacciani}, {Pellizzoni}, {Perotti}, {Picozza},
  {Pilia}, {Rapisarda}, {Rappoldi}, {Sabatini}, {Soffitta}, {Trifoglio},
  {Vercellone}, {Lucarelli}, {Santolamazza}, \& {Giommi}}]{Striani_etal-2011}
{Striani}, E., {Tavani}, M., {Piano}, G., {Donnarumma}, I., {Pucella}, G.,
  {Vittorini}, V., {Bulgarelli}, A., {Trois}, A., {Pittori}, C., {Verrecchia},
  F., {Costa}, E., {Weisskopf}, M., {Tennant}, A., {Argan}, A., {Barbiellini},
  G., {Caraveo}, P., {Cardillo}, M., {Cattaneo}, P.~W., {Chen}, A.~W., {De
  Paris}, G., {Del Monte}, E., {Di Cocco}, G., {Evangelista}, Y., {Ferrari},
  A., {Feroci}, M., {Fuschino}, F., {Galli}, M., {Gianotti}, F., {Giuliani},
  A., {Labanti}, C., {Lapshov}, I., {Lazzarotto}, F., {Longo}, F., {Marisaldi},
  M., {Mereghetti}, S., {Morselli}, A., {Pacciani}, L., {Pellizzoni}, A.,
  {Perotti}, F., {Picozza}, P., {Pilia}, M., {Rapisarda}, M., {Rappoldi}, A.,
  {Sabatini}, S., {Soffitta}, P., {Trifoglio}, M., {Vercellone}, S.,
  {Lucarelli}, F., {Santolamazza}, P., \& {Giommi}, P. 2011, Astrophys. J.
  Lett., 741, L5

\bibitem[{{Sturrock} \& {Aschwanden}(2012)}]{Sturrock_Aschwanden-2012}
{Sturrock}, P. \& {Aschwanden}, M.~J. 2012, Astrophys. J. Lett., 751, L32

\bibitem[{{Sturrock}(1971)}]{Sturrock-1971}
{Sturrock}, P.~A. 1971, Astrophys. J., 164, 529

\bibitem[{{Suttle} {et~al.}(2014){Suttle}, {Lebedev}, {Swadling},
  {Suzuki-Vidal}, {Burdiak}, {Bennett}, {Hare}, {Burgess}, {Clemens}, {Niasse},
  {Chittenden}, {Smith}, {Bland}, {Patankar}, \& {Stuart}}]{Suttle_etal-2014}
{Suttle}, L., {Lebedev}, S., {Swadling}, G., {Suzuki-Vidal}, F., {Burdiak}, G.,
  {Bennett}, M., {Hare}, J., {Burgess}, D., {Clemens}, A., {Niasse}, N.,
  {Chittenden}, J., {Smith}, R., {Bland}, S., {Patankar}, S., \& {Stuart}, N.
  2014, in APS Meeting Abstracts, 5007

\bibitem[{{Sweet}(1958)}]{Sweet-1958}
{Sweet}, P.~A. 1958, in IAU Symposium, Vol.~6, Electromagnetic Phenomena in
  Cosmical Physics, ed. B.~{Lehnert}, 123

\bibitem[{{Takahashi} \& {Ohsuga}(2013)}]{Takahashi_Ohsuga-2013}
{Takahashi}, H.~R. \& {Ohsuga}, K. 2013, Astrophys. J., 772, 127

\bibitem[{{Takahashi} \& {Ohsuga}(2015)}]{Takahashi_Ohsuga-2015}
{Takahashi}, H.~R. \& {Ohsuga}, K. 2015, in Thirteenth Marcel Grossmann
  Meeting: On Recent Developments in Theoretical and Experimental General
  Relativity, Astrophysics and Relativistic Field Theories, ed. K.~{Rosquist},
  2344--2345

\bibitem[{{Tavani} {et~al.}(2011){Tavani}, {Bulgarelli}, {Vittorini},
  {Pellizzoni}, {Striani}, {Caraveo}, {Weisskopf}, {Tennant}, {Pucella},
  {Trois}, {Costa}, {Evangelista}, {Pittori}, {Verrecchia}, {Del Monte},
  {Campana}, {Pilia}, {De Luca}, {Donnarumma}, {Horns}, {Ferrigno}, {Heinke},
  {Trifoglio}, {Gianotti}, {Vercellone}, {Argan}, {Barbiellini}, {Cattaneo},
  {Chen}, {Contessi}, {D'Ammando}, {DeParis}, {Di Cocco}, {Di Persio},
  {Feroci}, {Ferrari}, {Galli}, {Giuliani}, {Giusti}, {Labanti}, {Lapshov},
  {Lazzarotto}, {Lipari}, {Longo}, {Fuschino}, {Marisaldi}, {Mereghetti},
  {Morelli}, {Moretti}, {Morselli}, {Pacciani}, {Perotti}, {Piano}, {Picozza},
  {Prest}, {Rapisarda}, {Rappoldi}, {Rubini}, {Sabatini}, {Soffitta},
  {Vallazza}, {Zambra}, {Zanello}, {Lucarelli}, {Santolamazza}, {Giommi},
  {Salotti}, \& {Bignami}}]{Tavani_etal-2011}
{Tavani}, M., {Bulgarelli}, A., {Vittorini}, V., {Pellizzoni}, A., {Striani},
  E., {Caraveo}, P., {Weisskopf}, M.~C., {Tennant}, A., {Pucella}, G., {Trois},
  A., {Costa}, E., {Evangelista}, Y., {Pittori}, C., {Verrecchia}, F., {Del
  Monte}, E., {Campana}, R., {Pilia}, M., {De Luca}, A., {Donnarumma}, I.,
  {Horns}, D., {Ferrigno}, C., {Heinke}, C.~O., {Trifoglio}, M., {Gianotti},
  F., {Vercellone}, S., {Argan}, A., {Barbiellini}, G., {Cattaneo}, P.~W.,
  {Chen}, A.~W., {Contessi}, T., {D'Ammando}, F., {DeParis}, G., {Di Cocco},
  G., {Di Persio}, G., {Feroci}, M., {Ferrari}, A., {Galli}, M., {Giuliani},
  A., {Giusti}, M., {Labanti}, C., {Lapshov}, I., {Lazzarotto}, F., {Lipari},
  P., {Longo}, F., {Fuschino}, F., {Marisaldi}, M., {Mereghetti}, S.,
  {Morelli}, E., {Moretti}, E., {Morselli}, A., {Pacciani}, L., {Perotti}, F.,
  {Piano}, G., {Picozza}, P., {Prest}, M., {Rapisarda}, M., {Rappoldi}, A.,
  {Rubini}, A., {Sabatini}, S., {Soffitta}, P., {Vallazza}, E., {Zambra}, A.,
  {Zanello}, D., {Lucarelli}, F., {Santolamazza}, P., {Giommi}, P., {Salotti},
  L., \& {Bignami}, G.~F. 2011, Science, 331, 736

\bibitem[{{Tsuneta}(1996)}]{Tsuneta-1996}
{Tsuneta}, S. 1996, Astrophys. J., 456, 840

\bibitem[{{Uzdensky}(2006)}]{Uzdensky-2006}
{Uzdensky}, D.~A. 2006, ArXiv Astrophysics e-prints

\bibitem[{{Uzdensky}(2009)}]{Uzdensky-2009}
---. 2009, Physics of Plasmas, 16, 040702

\bibitem[{{Uzdensky}(2011)}]{Uzdensky-2011}
---. 2011, Space Sci. Rev., 160, 45

\bibitem[{{Uzdensky} {et~al.}(2011){Uzdensky}, {Cerutti}, \&
  {Begelman}}]{Uzdensky_etal-2011}
{Uzdensky}, D.~A., {Cerutti}, B., \& {Begelman}, M.~C. 2011, Astrophys. J.
  Lett., 737, L40

\bibitem[{{Uzdensky} \& {Goodman}(2008)}]{Uzdensky_Goodman-2008}
{Uzdensky}, D.~A. \& {Goodman}, J. 2008, Astrophys. J., 682, 608

\bibitem[{{Uzdensky} {et~al.}(2010){Uzdensky}, {Loureiro}, \&
  {Schekochihin}}]{Uzdensky_etal-2010}
{Uzdensky}, D.~A., {Loureiro}, N.~F., \& {Schekochihin}, A.~A. 2010, Physical
  Review Letters, 105, 235002

\bibitem[{{Uzdensky} \& {McKinney}(2011)}]{Uzdensky_McKinney-2011}
{Uzdensky}, D.~A. \& {McKinney}, J.~C. 2011, Physics of Plasmas, 18, 042105

\bibitem[{{Uzdensky} \& {Rightley}(2014)}]{Uzdensky_Rightley-2014}
{Uzdensky}, D.~A. \& {Rightley}, S. 2014, Reports on Progress in Physics, 77,
  036902

\bibitem[{{Uzdensky} \& {Spitkovsky}(2014)}]{Uzdensky_Spitkovsky-2014}
{Uzdensky}, D.~A. \& {Spitkovsky}, A. 2014, Astrophys. J., 780, 3

\bibitem[{{van Oss} {et~al.}(1993){van Oss}, {van den Oord}, \&
  {Kuperus}}]{van_Oss_etal-1993}
{van Oss}, R.~F., {van den Oord}, G.~H.~J., \& {Kuperus}, M. 1993, Astron. \&
  Astrophys., 270, 275

\bibitem[{{Vasyliunas}(1975)}]{Vasyliunas-1975}
{Vasyliunas}, V.~M. 1975, Rev. Geophys. and Space Phys., 13, 303

\bibitem[{{Werner} {et~al.}(2013){Werner}, {Begelman}, {Cerutti}, {Nalewajko},
  \& {Uzdensky}}]{Werner_etal-2013}
{Werner}, G., {Begelman}, M., {Cerutti}, B., {Nalewajko}, K., \& {Uzdensky}, D.
  2013, in APS Meeting Abstracts, 5003

\bibitem[{{Werner} {et~al.}(2014){Werner}, {Uzdensky}, {Cerutti}, {Nalewajko},
  \& {Begelman}}]{Werner_etal-2014}
{Werner}, G.~R., {Uzdensky}, D.~A., {Cerutti}, B., {Nalewajko}, K., \&
  {Begelman}, M.~C. 2014, ArXiv e-prints

\bibitem[{{Yamada} {et~al.}(2010){Yamada}, {Kulsrud}, \&
  {Ji}}]{Yamada_etal-2010}
{Yamada}, M., {Kulsrud}, R., \& {Ji}, H. 2010, Reviews of Modern Physics, 82,
  603

\bibitem[{{Yokoyama} {et~al.}(2001){Yokoyama}, {Akita}, {Morimoto}, {Inoue}, \&
  {Newmark}}]{Yokoyama_etal-2001}
{Yokoyama}, T., {Akita}, K., {Morimoto}, T., {Inoue}, K., \& {Newmark}, J.
  2001, Astrophys. J. Lett., 546, L69

\bibitem[{{Yokoyama} \& {Shibata}(1995)}]{Yokoyama_Shibata-1995}
{Yokoyama}, T. \& {Shibata}, K. 1995, Nature, 375, 42

\bibitem[{{Yu}(2012)}]{Yu-2012}
{Yu}, C. 2012, Astrophys. J., 757, 67

\bibitem[{{Yuan} {et~al.}(2011){Yuan}, {Yin}, {Wu}, {Bi}, {Liu}, \&
  {Zhang}}]{Yuan_etal-2011}
{Yuan}, Q., {Yin}, P.-F., {Wu}, X.-F., {Bi}, X.-J., {Liu}, S., \& {Zhang}, B.
  2011, Astrophys. J. Lett., 730, L15

\bibitem[{{Zenitani} \& {Hoshino}(2001)}]{Zenitani_Hoshino-2001}
{Zenitani}, S. \& {Hoshino}, M. 2001, Astrophys. J. Lett., 562, L63

\bibitem[{{Zenitani} \& {Hoshino}(2005)}]{Zenitani_Hoshino-2005}
---. 2005, Phys. Rev. Lett., 95, 095001

\bibitem[{{Zenitani} \& {Hoshino}(2007)}]{Zenitani_Hoshino-2007}
---. 2007, Astrophys. J., 670, 702

\bibitem[{{Zenitani} \& {Hoshino}(2008)}]{Zenitani_Hoshino-2008}
---. 2008, Astrophys. J., 677, 530

\bibitem[{{Zheleznyakov}(1977)}]{Zheleznyakov-1977}
{Zheleznyakov}, V.~V. 1977, {Electromagnetic waves in cosmic plasma. Generation
  and propagation.}

\bibitem[{{Zweibel} \& {Yamada}(2009)}]{Zweibel_Yamada-2009}
{Zweibel}, E.~G. \& {Yamada}, M. 2009, Annual Rev. Astron. \& Astrophys., 47,
  291

\end{thebibliography}


\end{document}